\documentclass[preprint]{elsarticle}

\usepackage{amsfonts}
\usepackage{url}

\begin{document}

\title{$SEIRS$ epidemiology model for the COVID-19 pandemy in the extreme case of no acquired immunity}

\author{J.M.~Ilnytskyi\corref{cor}}\ead{iln@icmp.lviv.ua}
\address{Institute for Condensed Matter Physics of Nat. Acad. Sci. of Ukraine, Lviv, Ukraine}
 
\date{\today}

\begin{abstract}
We consider the $SEIRS$ compartment epidemiology model suitable for predicting the evolution of the COVID-19 pandemy in the extreme limiting case of no acquired immunity. The disease-free and endemic fixed points are found and their stability is analysed. The expression for the basic reproduction ratio is obtained and discussed, emphasizing on its dependence on the model parameters. The threshold contact ratio is found which determines the possibility for a stable disease-free fixed point existence. Numeric solution for the pandemy evolution is also undertaken together with the approximate analytic solutions for the early stage of the disease spread as well as as for its decay after the rapid measures are undertaken. We analysed several possible scenarios for introducing and relaxing the quarantine measures. The cyclic ``quarantine on'' and ``quarantine off'' strategy at fixed identification and isolation ratios fail to reduce the lowering of the second and the consecutive waves, whereas this goal is possible to achieve if the flexible increase of the identification and isolation ratios is also involved.
\end{abstract}


\maketitle

\section{\label{I}Introduction and the model}

The COVID-19 pandemy, originated in 2019 and continuing its run at the moment of writing, is a \textit{serious multifaceted threat} to mankind. Until this day, there are 29 million registered cases and 924 thousands of deaths related to the illness directly or indirectly \cite{WHO_data}. With the absence of a vaccine, the only effective way of slowing it down, and, consequently, decreasing load on a local medical service undil medication is available, is introducing a set of quarantine measures. These are aimed on reduction of social contacts within a community and comprise closing down of travel roots, sport and cultural mass events, putting on restrictions on indoor factory works, shops, cafes, outdoor movement, etc. While effective at the initial phases of a pandemy, these measures put a huge strain on economy and normal functioning of a society. Therefore, one needs to search for compromised measures that, on one hand, enable normal functioning of a community, but on the other hand, keep the pandemy under control \cite{RojasVallejos2020,Adam2020,Eubank2020}. In this respect, development of suitable mathematical models that predict the realistic dynamics of a pandemy depending on the level of imposed or lifted up quarantine measures, plays a crucial role in developing a successful long-term strategy to defeat the pandemy.   

Modelling strategies related to COVID-19 are covered in detail in a number of recent reviews \cite{Park2020,Bai2020,Cheng2020}. A wide set of \textit{general epidemiology models}, $SIR$, $SEIR$, $SEIRU$, $SIRD$, $SLIAR$, $ARIMA$ and $SIDARTHE$ have been adapted for this purpose, as discussed in Ref.~\cite{Anirudh2020}. Most studies make use of $SIR$ or $SEIR$ epidemiology models modified to describe  the peculiarities of this disease. According to Carletti et al. \cite{Carletti2020}, the dynamics of the COVID-19 outbreak belongs to the simple universality class of the $SIR$ model and extensions thereof. There are also studies that use multi-group models to account for population heterogeneity, e.g. the multi-group $SEIR$ \cite{Nazarimehr2020} and the $SEIRA$ \cite{Contreras2020} ones.

From the practical point of view, the analysis and prediction of the pandemy outbreak in \textit{specific countries/regions} is of much interest and there are many papers devoted to such studies. In particular, the generalized $SEIR$ model was used to estimate the course of COVID-19 in Chile \cite{GuerreroNancuante2020} and in the UK \cite{Rawson2020}. A simple mathematic modeling approach, based on the analysis and fitting of available data, was used to track the outbreaks of COVID-19 in the US \cite{Tang2020a} and Brasil \cite{Tang2020b}. The real-life modelling for the dissemination of the COVID-19 in Italy is performed using the $SEIR$ model that includes a network of its 107 provinces with the use of precise data for the population mobility \cite{Gatto2020}. An extended $SEIRD$ model was used to describe the disease dynamics in Germany with the parameter values identified by matching the model output to the officially reported cases \cite{Gotz2020}. Two models of the $SEIR$ type are used to evaluate the role of the latency period of the infection in the dynamics of a COVID-19 epidemic in China \cite{Liu2020}. Besides the deterministic epidemiology models, the discrete-time Markov chain model was proposed that directly incorporates stochastic behavior and for which parameter estimation is straightforward from available data \cite{Chen2020}.

One of the characteristic (and rather unusual) feature of the COVID-19 pandemy is the abundance of \textit{asymptomatic cases} and cases with mild symptoms, for this and other peculiarities of the disease outbreak, see Ref.~\cite{Sun2020a}. Such infected individuals are unaware of being a host for the virus but may actively spread it around. In a real-life situation, a number of asymptomatic infected individuals are unknown, but their role in the pandemy dynamics can be estimated in mathematic modelling. Some modelling results predict that asymptomatic cases may lead to both fast outbreak and large outbreak size \cite{Sun2020b}. The role of asymptomatic infected individuals is also discussed in detail in Ref.~\cite{Chatterjee2020} in relation to the epidemic dissemination in India. The effect can be lessened by performing extensive and prompt identification of infected individuals by suitable testing protocols, as well as their consequent isolation Ref.~\cite{Carletti2020}.

At early stages of the COVID-19 dissemination over the globe (beginning of 2020), the main emphasis was put on its maximal possible slowing down, to prevent an overload of local medical services. Understanding the early transmission dynamics and evaluating the effectiveness of \textit{control measures} is crucial for sustaining the disease dissemination in new areas \cite{Kucharski2020,Rahimi2020}. One of the first measures, introduced both intra- and internationally, were travel restrictions, as these should prevent the dissemination of a virus on a non-local scale. The precise effects of these measures are, however, unknown, therefore, a number of studies addressed them explicitly. 

In particular, the simulations of a global network mobility model based on air traffic in Europe \cite{Linka2020} show that mobility networks of air travel can predict the emerging global diffusion pattern of a pandemic at the early stages of the outbreak. The effects of travel restrictions, as well as social distancing, hospitalization, quarantine and hygiene measures on short-and-long term dynamics of the COVID-19 is examined and combined with data in South Africa during March-May 2020 \cite{Mushayabasa2020}.  By combining a global network mobility model with a local epidemiology model the simulations predict the outbreak dynamics of COVID-19 across Europe and the results suggest that an unconstrained mobility would have significantly accelerated the spreading of COVID-19, especially in Central Europe, Spain, and France \cite{Linka2020}. The impact of the other measures, such as school closures, physical distancing, shielding of people aged 70 years or older, and self-isolation of symptomatic cases, were investigated for the case of the UK in Ref.~\cite{Davies2020}. It was suggested that to end the COVID-19 epidemic, social distancing and wearing masks are absolutely crucial, along with the policy of reducing the transmission period by finding and isolating patients as quickly as possible through the efforts of the quarantine authorities \cite{Choi2020}.

The public-health policies to keep pandemic under control are also considered in terms of the $SEIRA$ model in Ref.~\cite{Contreras2020}. The outbreak dynamics of COVID-19 in China and the United States was quantified by using a global network model with a local epidemic $SEIR$ model. When postulating that the latent and infectious periods are disease-specific and the contact period is behavior-specific, this network model predicts that without the massive political mitigation strategies the United States would have faced a basic reproduction number of about 5.3 and a nationwide pandemy peak \cite{Peirlinck2020}. This was found to be a realistic prognosis in the following months after the publication date (May 2020). Scenarios of different containment measures and their impact were analyzed using the $SEIR$ on a network of 107 provinces in Italy with known population mobility \cite{Gatto2020}. Results suggest that the mobility restrictions and reduction of human-to-human interactions have reduced transmission by about 45\%, resulting in the conclusion that verifiable evidence exists to support the planning of emergency measures. Data collected over disease cases in Hubei province were analysed using the Bayesian approach \cite{Zhang2020}. The estimates suggest an early peak of infectiousness, with possible transmission before the onset of symptoms. Obtained results also indicate that, as the epidemic progressed, infectious individuals should be isolated quicker, to shorten the window of transmission in the community. All these studies indicate that strict containment measures, movement restrictions, and increased awareness of the population might have contributed to interrupt local transmission of COVID-19 \cite{Zhang2020,Choi2020}.

At further stages of the COVID-19 pandemy, it became quite clear that severe quarantine measures (while mostly effective in slowing a pandemy down and preventing health systems overload) yield significant adverse economic consequences. As the COVID-19 pandemy seems to be a long-term affair, one should seek for some compromises solutions of \textit{safe ways for relaxing and partial lifting the quarantine measures}. As suggested in Ref.~\cite{Chowdhury2020}, dynamic restrictions, with intervals of relaxed social distancing, may provide a good option. The authors of this study aimed on finding an ideal frequency and duration of the periods of quarantine restrictions, mitigation and relaxation by using the multivariate prediction model based on up-to-date transmission and clinical parameters in 16 countries. It was found that dynamic cycles of 50-day quarantine mitigation followed by a 30-day relaxation reduced transmission, but cannot reduce hospitalizations below required limits. This, however, can be achieved by employing cycles of 50-day suppression followed by a 30-day relaxation. This multi-country analysis predicts that a combination of quarantine measures and their relaxation can be employed as the effective strategy for COVID-19 pandemic control. Fusing models from epidemiology and network science, Block et al. \cite{Block2020} show how to ease lockdown and slow infection dissemination by strategic modification of people's contacts. Using the approach of social network, they evaluated the effectiveness of three distancing strategies such as: limiting interaction to a few repeated contacts (social bubbles), looking for similarity across contacts, and strengthening communities via triadic strategies. It was demonstrated that a strategic social network-based reduction of contact strongly enhances the effectiveness of social distancing measures while keeping risks lower and this provides an evidence for effective social distancing mitigating negative consequences of social isolation \cite{Block2020}.

Adapted $SEIR$ model was used to investigate the efficacy of two potential lockdown release strategies, focusing on the UK population as a test case \cite{Rawson2020}. Ending quarantine for the entire population simultaneously was found as a high-risk strategy, and that a gradual re-integration approach would be more reliable. However, lockdown should not be relaxed until the number of new daily confirmed cases reaches a sufficiently low threshold. Using optimization methods with adapted $SEIR$ model, it was found that the optimal strategy is to release approximately half the population 2-4 weeks from the end of an initial infection peak, then wait another 3-4 months to allow for a second peak before releasing everyone else. The extreme ``on-off'' strategy of releasing everyone, but re-establishing lockdown if the number of cases raises again, is found to be too risky. The worst-case scenario of a gradual release is found to be more manageable than the worst-case scenario that used the threshold based on-off strategy \cite{Rawson2020}.

Lop\'{e}z et al. explore different post-confinement scenarios by using a stochastic modified $SEIR$ model that accounts for the dissemination of infection during the latent period and also incorporates time-decaying effects due to potential loss of acquired immunity, people's increasing awareness of social distancing and the use of non-pharmaceutical interventions. Results being obtained suggest that lockdowns should remain in place for at least 60 days to prevent epidemic growth, as well as a potentially larger second wave of COVID-19 cases occurring within months. The best-case scenario should also gradually incorporate workers in a daily proportion at most 50\% higher than during the confinement period. Decaying immunity and particularly awareness and behaviour have 99\% significant effects on both the current wave of infection and on preventing COVID-19 reemergence. Social distancing and individual non-pharmaceutical interventions could potentially remove the need for lockdowns \cite{Lopez2020}.

One can summarize the findings outlined above by saying that both actuality and complexity of the COVID-19 pandemy call for employment of the modelling of various types that take into account COVID-19 characteristic features. From these we emphasize: (i) \textit{abundance of asymptomatic infected individuals}; (ii) \textit{absence of effective medication}, and (iii) \textit{loss of acquired immunity and virus mutation}. While the (i) and (ii) are quite unambiguous, the feature (iii) calls for some discussion. During the first months of the COVID-19 outbreak, it was widely assumed that after recovery one acquires immunity and it will be safe for such person to perform a social work in the places with a higher risk to be infected. This belief led to a widely discussed concept of ``immunity passports'' \cite{WHO_imm_pass}. It, however, turned out that such immunity may last 3-4 months only, as shown by immunodiagnostic tests \cite{Edridge2020,Ibarrondo2020,Springer2020}, resulting in reported cases of reinfections \cite{Tillett2020,Vrieze2020,Edridge2020,Ibarrondo2020,Springer2020,Forbes}. Reported virus mutations \cite{Mallapaty2020,Hou2020,Goodman,Terry} contribute an additional factor to the reinfection scenario. In this case the classical concept of recovered and immunized individual is valid within the time scale of 3-4 months only. There are, however, indications that the duration of pandemy may be as much as several years and, on this time scale, one needs to account for a loss of acquired immunity \cite{Lopez2020} and of mutation of a virus. In this respect, it makes sense to consider the extreme case of \textit{complete lack of the acquired immunity}, as the most harsh scenario. This is done in this study.

\begin{figure}[!ht]
\begin{center}
\includegraphics[clip,width=8cm,angle=0]{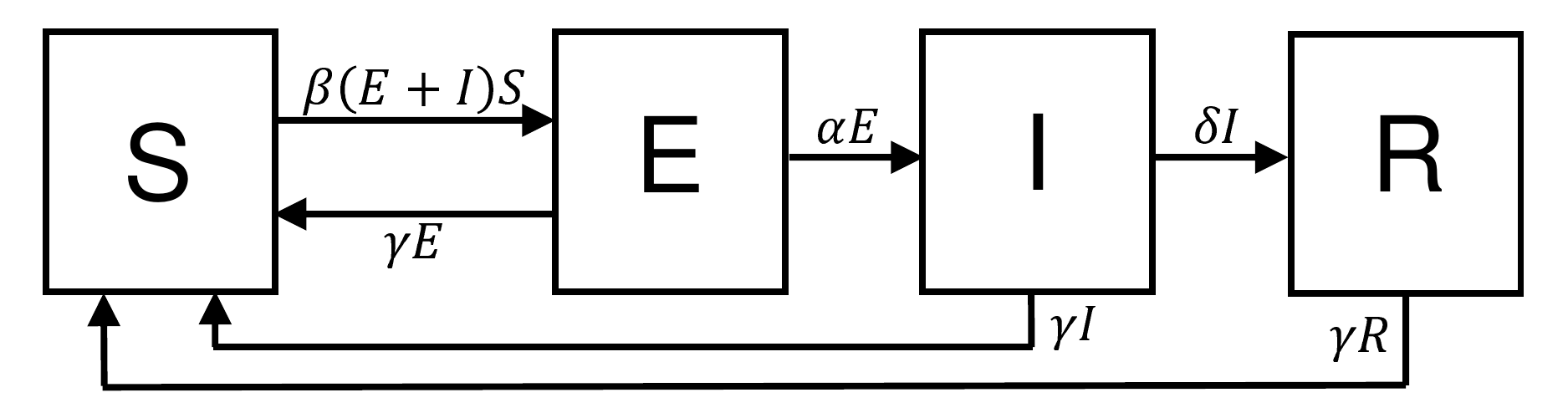}
\caption{\label{Model}The $SEIRS$ epidemiology model modified for the COVID-19 dissemination.}
\end{center}
\end{figure}

The $SEIRS$ compartmental model, suggested in this study and shown schematically in Fig.~\ref{Model}, contains four groups, marked via their respective fractions of susceptible $S$, unidentified infected $E$, identified infected $I$ and isolated $R$ individuals. Group $E$ contains individuals within incubation period of the disease, asymptomatic patients and these with mild symptoms, they are the susceptible individuals from group $S$ infected with the contact rate $\beta$. The $E$ individuals y are transfered further to group $I$ of identified infected individuals, those tested positive via PCR or other tests, with the identification rate $\alpha$. Then, identified infected individuals are isolated (in hospital or in home) by transferring them to the group $R$ with the isolation rate $\delta$. Therefore, the $R$ group contains isolated infected individuals rather then recovered ones, as in classical $SEIR$ model. Infected individuals that are not isolated yet, those from $E$ and $I$ groups, are assumed to be contagious. All infected individuals, from the groups $E$, $I$ and $R$, regardless of their identification and isolation status, recover with the same rate $\gamma$ (accounts for lack of medication), and are transfered back to the group $S$, where they can be infected again (accounts for complete lack of immunity). To simplify the model, the birth and death events are not taken into account, assuming that the fraction of death cases with respect to the total population is small.

The corresponding set of differential equations has the following form:
\begin{eqnarray}
\dot{S} & = & -\beta(E+I)S + \gamma (1-S) \label{dSdt}\\
\dot{E} & = & \beta(E+I)S - (\gamma + \alpha) E \label{dEdt}\\
\dot{I} & = & \alpha E - (\gamma + \delta) I \label{dIdt}\\
\dot{R} & = & \delta I - \gamma R \label{dRdt}
\end{eqnarray}
with an additional constraint of $S+E+I+R=1$, which is already used to simplify Eq.~(\ref{dSdt}).

The purpose of this study is to examine the stationary states and the time evolution of such $SEIRS$ model with the emphasis put on the influence of the quarantine measures and their relaxation on pandemy dynamics. The study is of a general type with no direct link to particular country/region or statistical data of any sort. We, therefore, concentrate on features and tendencies as predicted by this model and not on practical recommendations that can be used straightaway. Section \ref{II} contains analysis of the stationary states (fixed points) for the model and analysis of their stability; in section \ref{III} we discuss early-time spread and the decay dynamics of the disease dissemination by combining numerical and approximate analytic tools; in section \ref{IV} we consider the effects for quarantine measures and their relaxation on the dynamics of the COVID-19 pandemy, especially on the height of the second wave of the disease, section \ref{V} contains conclusions.  

\section{Fixed points and their stability}\label {II}

The stationary state (fixed points) for the $SEIRS$ model is given as the solution of the equation set:
\begin{eqnarray}
&&-\beta(E+I)S+ \gamma (1-S) = 0\label{S1}\\
&&\beta(E+I)S - (\gamma + \alpha) E = 0\label{S2}\\
&&\alpha E - (\gamma + \delta) I = 0\label{S3}\\
&&\delta I - \gamma R = 0\label{S4}\\
&&S+E+I+R=1\label{S5}
\end{eqnarray}
Eqs.~(\ref{S3}) and (\ref{S4}) allow to express $E$ and $R$ via $I$

\begin{equation}
 E=\frac{\gamma+\delta}{\alpha}I,\hspace{1em} R=\frac{\delta}{\gamma}I
\end{equation}
Substituting the first one into Eq.~(\ref{S1}) and combining it with Eq.~(\ref{S5}) one obtains the set of equations for $S$ and $I$ in a stationary state
\begin{eqnarray}
&&-\beta(\gamma+\delta+\alpha)IS+ (\gamma+\alpha)(\gamma+\delta)I = 0\label{SE1}\\
&&S + \frac{(\gamma+\delta)(\gamma+\alpha)}{\alpha\gamma}I = 0\label{SE2}
\end{eqnarray}
There are two fixed points. The disease-free (DF) fixed point is
\begin{equation}
S^\dagger=1,\hspace{1em} E^\dagger=I^\dagger=R^\dagger=0\label{dfr_stat}
\end{equation}
Then, assuming $I\neq 0$, we obtain the other, endemic (EN), fixed point
\begin{eqnarray}
S^*&=&\max\left\{\frac{(\gamma+\alpha)(\gamma+\delta)}{\beta(\gamma+\alpha+\delta)},1\right\}\label{Sstat}\\
E^*&=&\frac{\gamma}{\gamma+\alpha}(1-S^*)\label{Estat}\\
I^*&=&\frac{\alpha\gamma}{(\gamma+\alpha)(\gamma+\delta)}(1-S^*)\label{Istat}\\
R^*&=&\frac{\alpha\delta}{(\gamma+\alpha)(\gamma+\delta)}(1-S^*)\label{Rstat}
\end{eqnarray}
The crossover between two fixed points occurs at $S^*=1$. As far as here the disease dies out, one can associate the expression
\begin{equation}\label{Rnumb}
\frac{\beta(\gamma+\alpha+\delta)}{(\gamma+\alpha)(\gamma+\delta)}=R_0
\end{equation} 
with the basic reproductive number $R_0$. It is symmetric with respect to the $\alpha$ and $\delta$ and reduces to the $\beta/\gamma$ when $\alpha=0$ (the $SES$ model) or $\delta=0$ (the $SE'S$ model, where $E'=E+I$). At fixed $\gamma$, $R_0$ is the function of three independent model parameters, the contact $\beta$, identification $\alpha$, and isolation $\delta$ rates. Full differential $dR_0$ of the basic reproductive number can be found easily and reads
\begin{equation}\label{dRnumb}
\frac{1}{R_0}dR_0=\frac{1}{\beta}d\beta - \frac{1}{\gamma+\alpha+\delta}\left(\frac{\delta}{\gamma+\alpha}d\alpha+\frac{\alpha}{\gamma+\delta}d\delta\right).
\end{equation} 
It provides quantitative means on how exactly the infinitesimal decrease $d\beta<0$ of the contact rate and the infinitesimal increase of either identification $d\alpha>0$ or isolation $d\delta>0$ rates affect the change $dR_0$ of the basic reproductive number. In practical terms, both options have associated financial burden: quarantine related economy losses for the first option; and the cost of extensive coverage of population by medical tests (e.g. PCR) and the isolation costs (control of home isolation and hospital care) for the second one. Exact monetary expenses for both options depend on many economic and social details that are country dependent, and their estimation are beyond this study. However, if known, these can be used to obtain practical estimates for the economic efficiency of both options of reduction of the basic reproductive number using Eq.~(\ref{dRnumb}). Another point to mention is, that the relative, $\frac{d\beta}{\beta}$, and not the absolute, $d\beta$ change of the contact rate enters Eq.~(\ref{dRnumb}). This shows that more strict quarantine measures (larger magnitude of $d\beta$) are needed in the countries with initially high contact rates $\beta$, as compared to those with a low contact rate, to achieve the same reduction of the basic reproductive number $R_0$.

One may find certain ``equivalence'' between the effects of reduction of $\beta$ and on the increase of two other rates, $\alpha$ and $\delta$, when imposing the the condition that the basic reproduction number is unchanged, $dR_0=0$. As follows from Eq.~(\ref{dRnumb}), in this case
\begin{equation}\label{dRnumb_const}
\frac{1}{\beta}d\beta = \frac{1}{\gamma+\alpha+\delta}\left(\frac{\delta}{\gamma+\alpha}d\alpha+\frac{\alpha}{\gamma+\delta}d\delta\right)
\end{equation} 
In a special case of $\alpha=\delta$, one has $\frac{d\beta}{\beta}=\frac{2}{(2+\gamma/\alpha)(1+\gamma/\alpha)}\frac{d\alpha}{\alpha}$, which simplifies further to $\frac{d\beta}{\beta}=\frac{d\alpha}{\alpha}$ if $\gamma\ll\alpha$. In this oversimplified case, the relative infinitesimal change of the contact rate is balanced by exactly the same relative change of the identification/isolation rate to keep the same value of the basic reproductive number $R_0$.

\begin{figure}[!ht]
\begin{center}
\includegraphics[clip,width=12cm,angle=270]{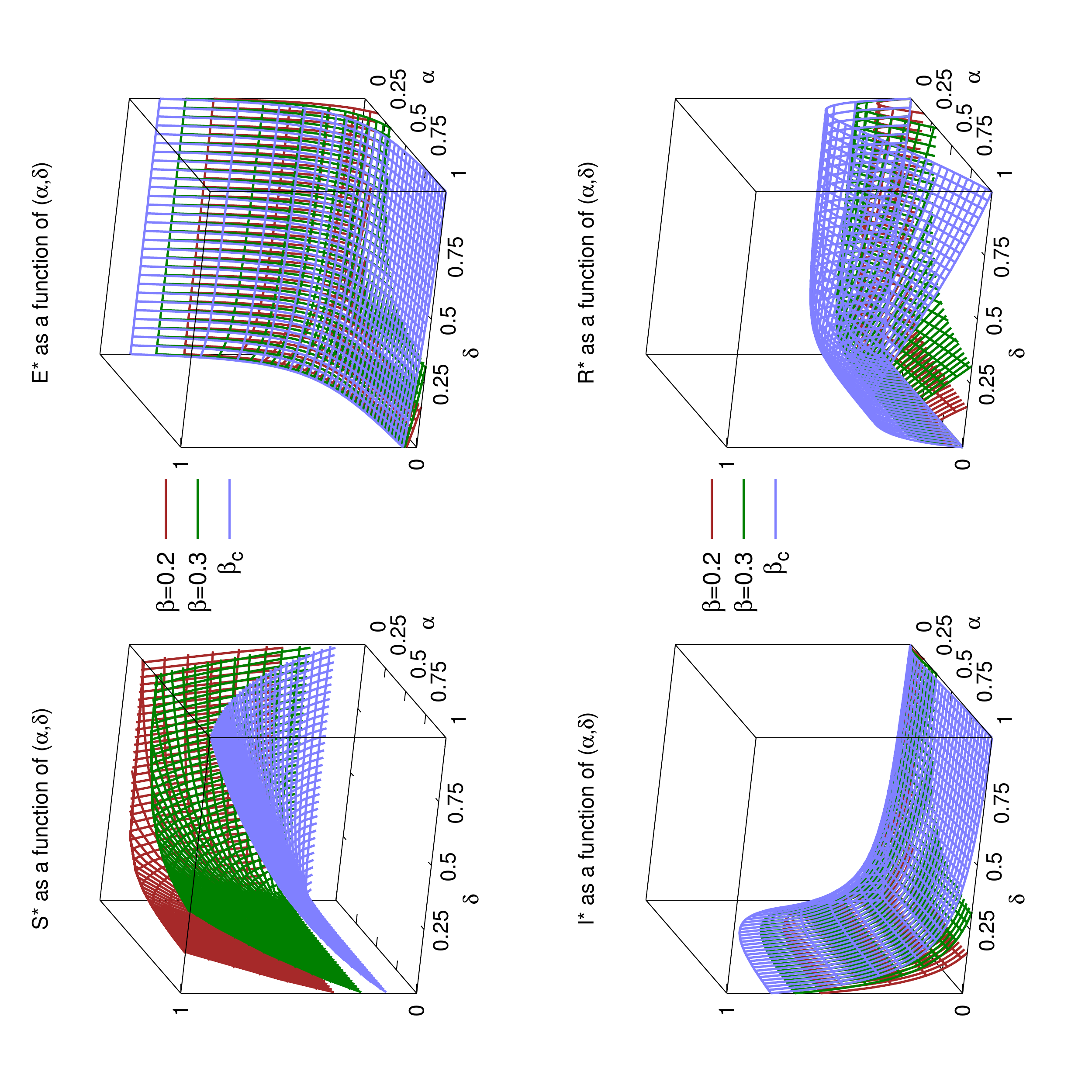}
\caption{\label{E_EI_R_EIR_3D}The $3D$ plots for the fractions $S^*$, $E^*$, $I^*$ and $R^*$  as the functions of $(\alpha,\delta)$. A series of contact rates $\beta$ from $0.1$ to the critical value $\beta_c$ (\ref{betac}) are considered, curing rate is fixed at $\gamma=1/14$.}
\end{center}
\end{figure}
Let us find the regions in the parameter space $\{\alpha,\delta,\beta,\gamma\}$, where the DF and EN fixed points exist, termed thereafter as the DF and EN regions, respectively. These are shown in Fig.~\ref{E_EI_R_EIR_3D} as surface plots for all the fractions with respect to $\alpha$ and $\delta$ build at a range of contact rates $\beta$. The curing rate is fixed at $\gamma=1/14$ throughout this study assuming the average curing period of $14$ days. The plots are symmetric with respect to $\alpha$ and $\delta$, as follows from the Eqs.~({\ref{Sstat})-(\ref{Rstat}). The plot for $S^*$ indicates that $S^*<1$ (the EN fixed point) turns into the $S^\dagger=1$ (the DF fixed point) at the crossover curve defined by
\begin{equation}
\frac{(\gamma+\alpha)(\gamma+\delta)}{\beta(\gamma+\alpha+\delta)} = 1.\label{crossover}\\
\end{equation}
Therefore, the DF region point spans in between this curve and the $\alpha=\delta=1$ point.
\begin{figure}[!ht]
\begin{center}
\includegraphics[clip,width=5cm,angle=270]{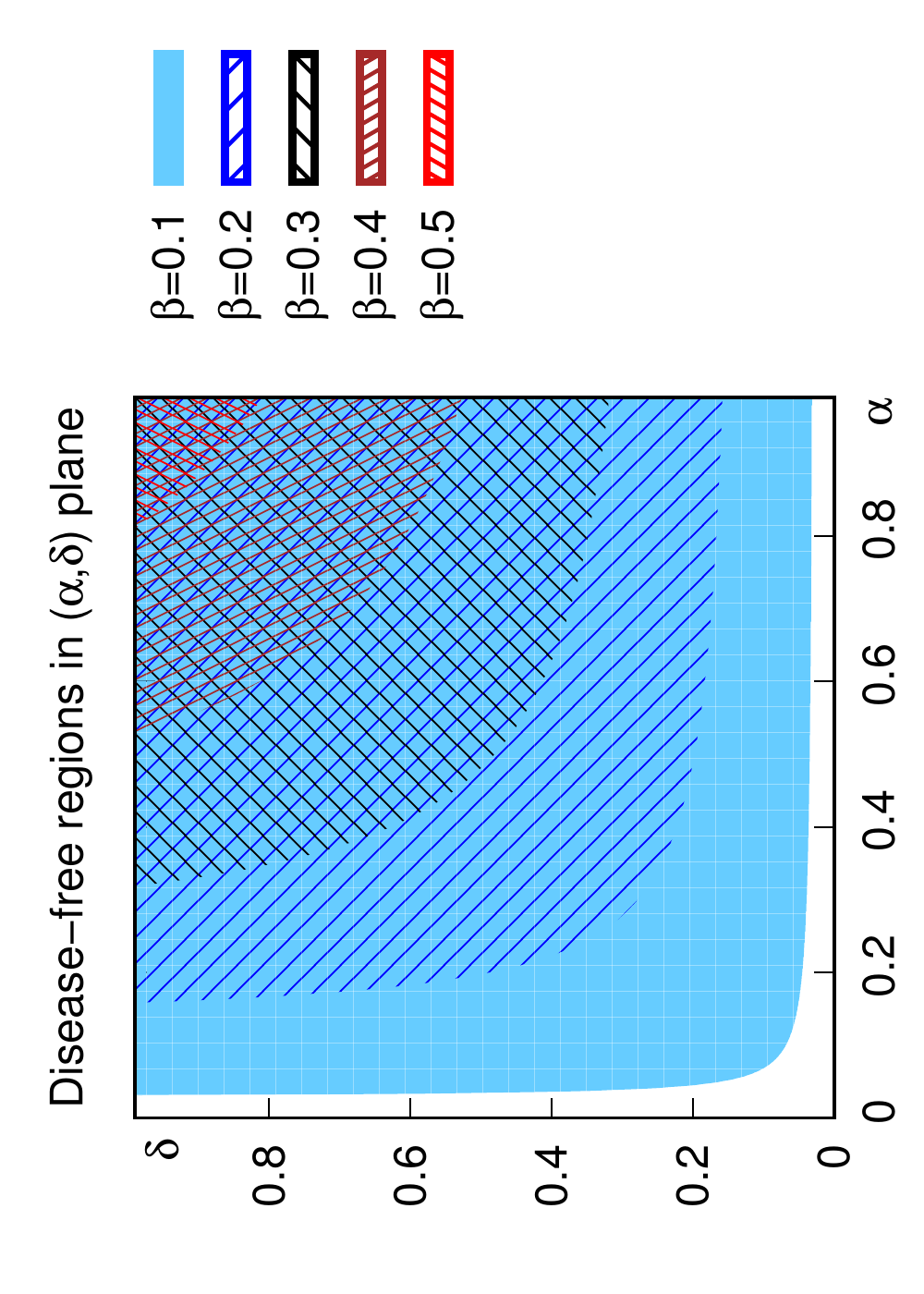}
\caption{\label{Dis_free}DF regions in $(\alpha,\delta)$ space.}
\end{center}
\end{figure}
The area of the DF region shrinks with the increase of $\beta$, as shown by dashed regions in Fig.~\ref{Dis_free}, and disappears at the critical value 
\begin{equation}\label{betac}
\beta_c=\frac{(\gamma+1)^2}{\gamma+2},
\end{equation} 
for the contact rate. At $\gamma=1/14$, this value is $\beta_c\approx 0.554$. At the higher contact rate, $\beta\geq\beta_c$, a DF region disappears. The other plots, for $E^*$, $I^*$ and $R^*$, in Fig.~\ref{E_EI_R_EIR_3D} show the redistribution of the individuals between respective compartments following changes in the identification $\alpha$ and the isolation $\delta$ rates. These all cross the $z=0$ plane at the same crossover curve (\ref{crossover}) and are equal to zero in the DF region. 

Now we will proceed to the question of stability of the fixed points in their respective regions. This is performed via linear stability analysis based on the Jacobian matrix for the set of equations (\ref{dSdt})-(\ref{dRdt})
\begin{equation}
\mathbf{J}=\left(
\begin{array}{cccc}
-\beta(E+I)-\gamma & -\beta S & -\beta S & 0\\
\beta(E+I) & \beta S-(\gamma+\alpha) & \beta S & 0\\
0 & \alpha & -(\gamma+\beta) & 0\\
0 & 0 & \delta & -\gamma
\end{array}
\right)
\end{equation}
The characteristic equation $F(\lambda)=\det(\mathbf{J-\lambda\mathbf{1}})=0$ (here $\mathbf{1}$ is the unit matrix) for the eigenvalues $\lambda_i$ of this matrix can be split into
\begin{equation}
F(\lambda)=F_{R}(\lambda)F_{SEI}(\lambda)=0,
\end{equation}
where $F_{R}(\lambda)$ originates from the last row of $\mathbf{J}$ and $F_{SEI}(\lambda)$ -- from the first three rows
\begin{eqnarray}
F_{R}(\lambda) &=& \lambda+\gamma\label{FR_term}\\ 
F_{SEI}(\lambda) &=& \lambda^3+\big[(\gamma+\alpha)+(\gamma+\delta)+(\gamma-\beta(S-E-I))\big]\lambda^2\nonumber\\
&+&\big[(\gamma+\alpha)(\gamma+\delta)+((\gamma+\alpha)+(\gamma+\delta))(\gamma-\beta(S-E-I))\big]\lambda\nonumber\\
&+&(\gamma+\alpha)(\gamma+\delta)(\gamma+\beta(E+I))-\gamma\beta(\gamma+\alpha+\delta S) \label{char_eq}
\end{eqnarray}
For the fixed point to be stable, the real parts of all eigenvalues $\lambda_i$ should be negative. The first eigenvalue $\lambda_1=-\gamma$ is obtained trivially from the $F_{R}(\lambda)=0$ equation and is the same and negative for both fixed points since curing rate $\gamma$ is always positive. The rest eigenvalues, $\lambda_2$, $\lambda_3$ and $\lambda_4$, are the solutions of the $F_{SEI}(\lambda)=0$ equation and differ for the cases of the DF and EN fixed points.

\begin{figure}[!ht]
\begin{center}
\includegraphics[clip,width=6cm,angle=270]{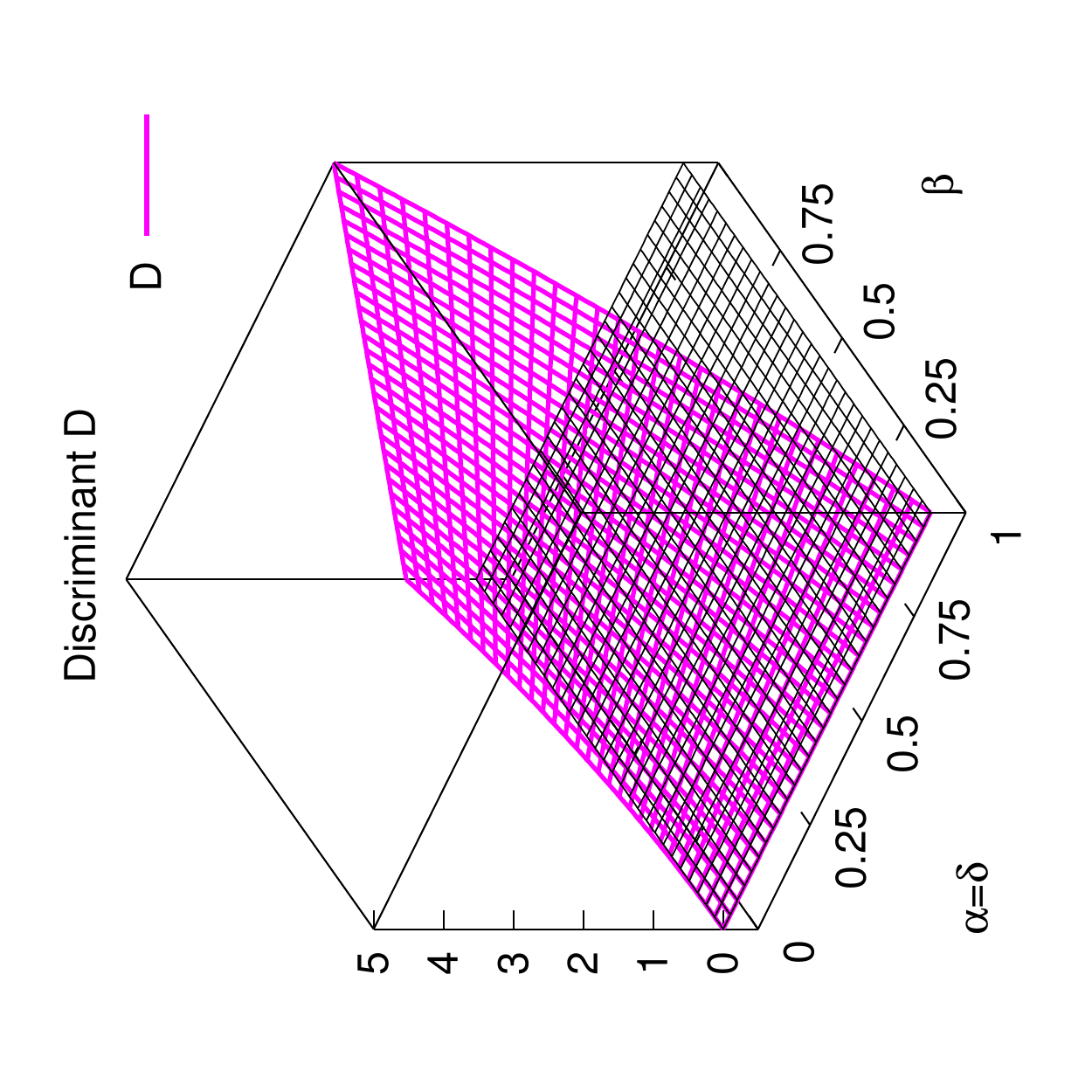}
\includegraphics[clip,width=6cm,angle=270]{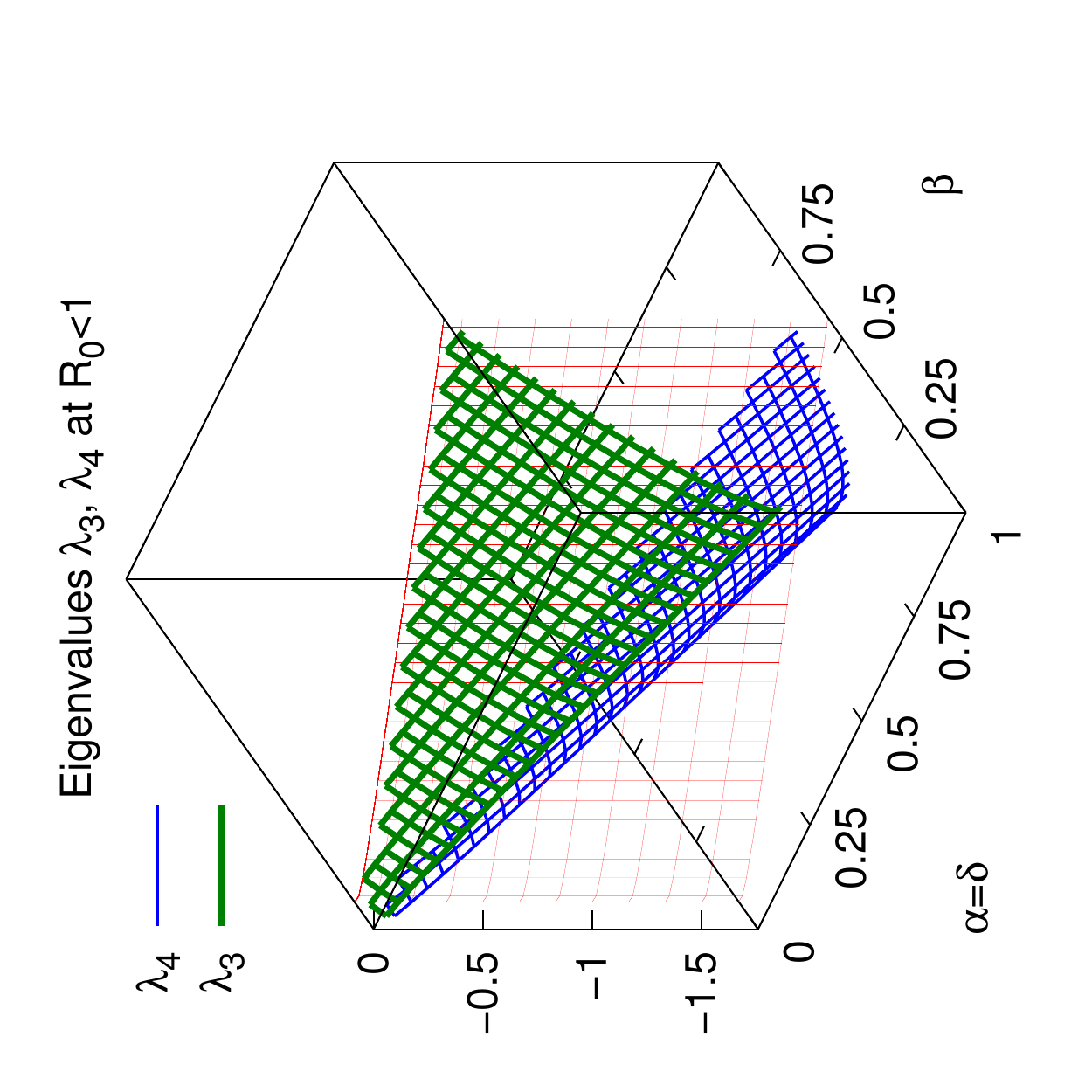}
\caption{\label{Disfree_D_eigen} Left side: discriminant $D$ of the quadratic equation $F_{DF}(\lambda)=0$ for the eigenvalues $\lambda_3$ and $\lambda_4$ at the DF fixed point. Right side: $\lambda_3$ and $\lambda_4$ in the DF region ($R_0<1$). The crossover surface is shown in red.}
\end{center}
\end{figure}
In particular, for the DF fixed point (\ref{dfr_stat}), the characteristic equation $F_{SEI}(\lambda)=0$ for $\lambda_2$, $\lambda_3$ and $\lambda_4$,  simplifies to
\begin{eqnarray}
F_{SEI}(\lambda)&=&(\lambda+\gamma)F_{DF}(\lambda)=0,~~\mathrm{where}\label{DF_fact}\\
F_{DF}(\lambda)&=&\lambda^2+\big[(\gamma+\alpha)+(\gamma+\delta)-\beta\big]\lambda\nonumber\\
                          &+&(\gamma+\alpha)(\gamma+\delta)-\beta(\gamma+\alpha+\delta).\label{DFEq}
\end{eqnarray}
Hence, one of the eigenvalues, $\lambda_2=-\gamma$, is found easily from Eq.~(\ref{DF_fact}) and is equal to $\lambda_1$ and always negative. Two other eigenvalues, $\lambda_3$ and $\lambda_4$, are solutions of the $F_{DF}(\lambda)=0$ equation. The signs of their real parts of are examined graphically for the simplified case of $\alpha=\delta$ to enable visualisation of the analysis via $3D$ plots. The results are shown in Fig.~\ref{Disfree_D_eigen}. The plot on the left shows the discriminant $D$ of the square equation $F_{DF}(\lambda)=0$, which is found to be positive at all $\beta$ and $\alpha=\delta$ if both rates are contained within the $[0:1]$ interval. Therefore, both $\lambda_3$ and $\lambda_4$ are real. The plot on the right shows $\lambda_3$ and $\lambda_4$ within the DF region ($R_0<1$) delimited by a crossover surface shown in red, which is obtained by translation of the crossover curve (\ref{crossover}) along the $Z$-axis. Both $\lambda_3$ and $\lambda_4$ are negative within the DF region indicating stability of the DF fixed point here. This is found to hold for the more general, $\alpha\neq\delta$, case (not shown here).

\begin{figure}[!ht]
\begin{center}
\includegraphics[clip,width=6cm,angle=270]{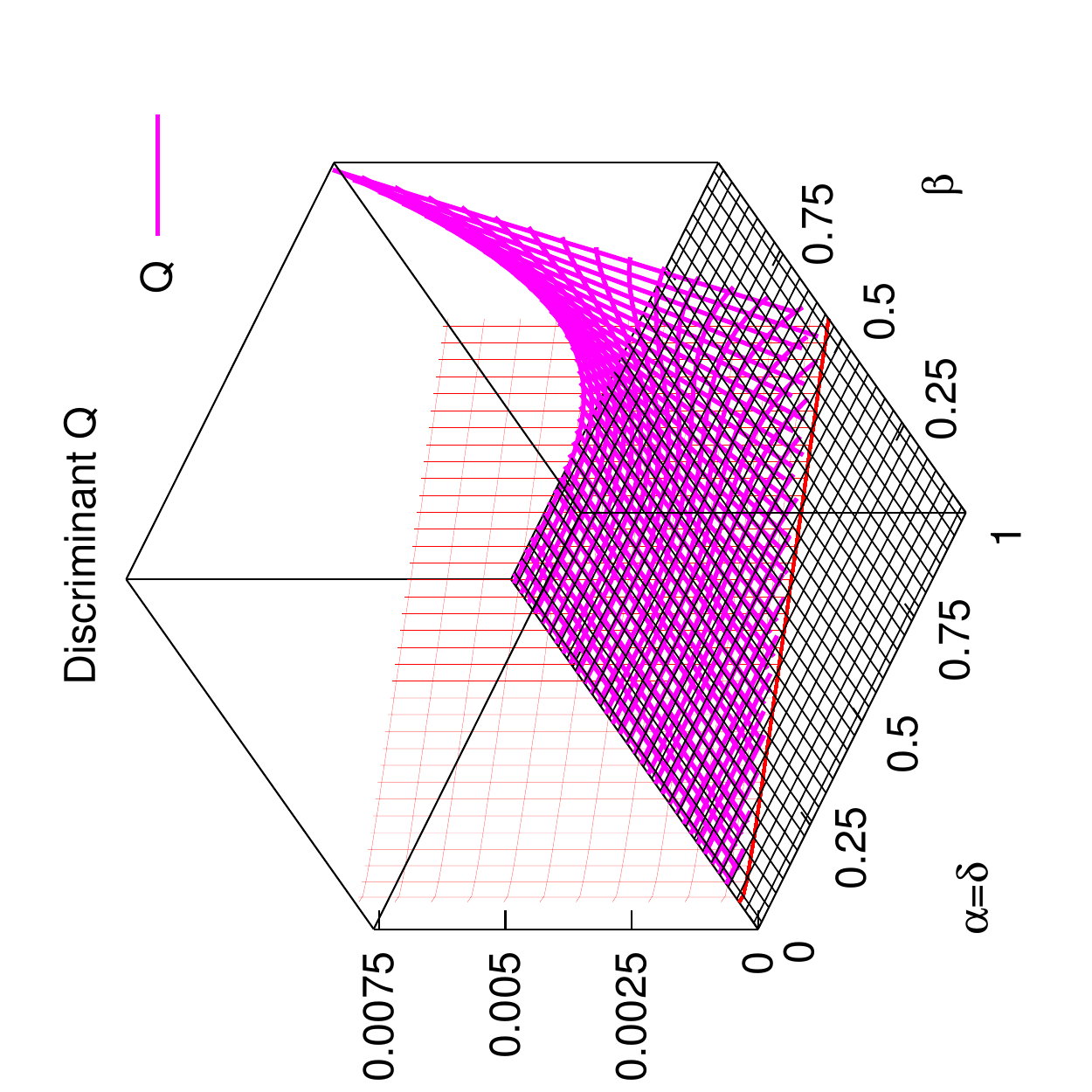}
\includegraphics[clip,width=6cm,angle=270]{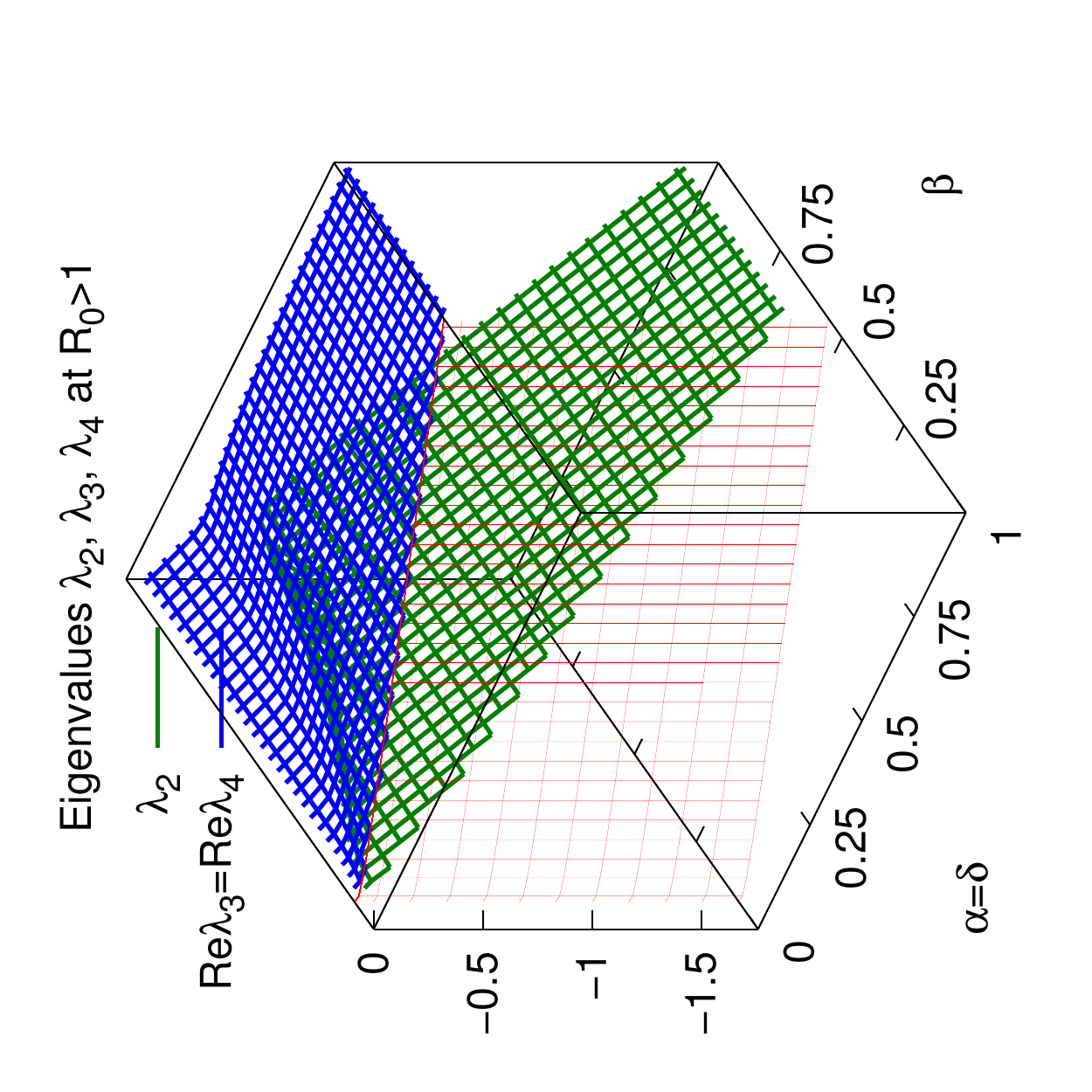}
\caption{\label{Endem_Q_eigen}Left side: Discriminant $Q$ of the cubic equation $F_{SEI}(\lambda)=0$ for the eigenvalues $\lambda_2$, $\lambda_3$ and $\lambda_4$ at the EN fixed point. Right side: $\lambda_2$, $\lambda_3$ and $\lambda_4$ in the EN region ($R_0>1$). The crossover surface is shown in red.}
\end{center}
\end{figure}

For the case of the EN fixed point (\ref{Sstat})-(\ref{Rstat}), the characteristic equation $F_{SEI}(\lambda)=0$ reads  
\begin{eqnarray}
F_{SEI}(\lambda)&=&F_{EN}(\lambda)=0,~~\mathrm{where}\nonumber\\
F_{EN}(\lambda)&=&\lambda^3+\big[(\gamma+\alpha)+(\gamma+\delta)+(\gamma/S^*-\beta S^*)\big]\lambda^2\nonumber\\
&+&\big[(\gamma+\alpha)(\gamma+\delta)+((\gamma+\alpha)+(\gamma+\delta))(\gamma/S^*-\beta S^*)\big]\lambda\label{ENEq}\\
&+&\gamma\big[\beta(\gamma+\alpha+\delta)-(\gamma+\alpha)(\gamma+\delta)\big]\nonumber\\
&\equiv& \lambda^3 + a\lambda^2+b\lambda+c = 0, \label{abc}
\end{eqnarray}
where $a$, $b$ and $c$ are the coefficients next to the powers of $\lambda$, introduced for the sake of brevity. This equation can be rewritten in reduced form
\begin{equation}
\mu^3 + p\mu +q = 0, \label{Eqmu}
\end{equation}
where
\begin{equation}
p=-\frac{a^2}{3}+b,~~ q=\left(\frac{a}{3}\right)^3,~~ \lambda = \mu-\frac{a}{3}.\label{pqlam}
\end{equation}
This enables to use the Cardano formula \cite{korn2013mathematical}, in which case the solutions, $\mu_2$, $\mu_3$ and $\mu_4$, depend on the sign of the discriminant
\begin{equation}
Q=\left(\frac{p}{3}\right)^3+\left(\frac{q}{2}\right)^2\label{Q}
\end{equation}

The sign of $Q$ is examined graphically for the simplified case of $\alpha=\delta$ and is found to be positive within the whole EN region, as shown in Fig.~\ref{Endem_Q_eigen}, on the left. In this case, one of the solutions, $\mu_2$ is real, whereas two others, $\mu_3$ and $\mu_4$ are complex, and we will be interested in their real parts only. The respective expressions are \cite{korn2013mathematical}
\begin{equation}
\mu_2=A+B,~~ \mathrm{Re}(\mu_3)=\mathrm{Re}(\mu_4)=\frac{A+B}{2}-\frac{p}{3}
\end{equation}
where
\begin{equation}
A=-\frac{q}{2}+\sqrt[3]{Q},~~ B=-\frac{q}{2}-\sqrt[3]{Q}
\end{equation}
The real parts of the eigenvalues $\lambda_i$ can be found from here according to Eq.~(\ref{pqlam}) and these are shown within the EN region in the right plot of Fig.~\ref{Endem_Q_eigen}. One can see that $\lambda_2$ and the real parts of both $\lambda_3$ and $\lambda_4$ are all negative, hence the EN fixed point is stable in the EN region.

To conclude this section, we found two fixed points, the disease-free and the endemic one and the respective regions of parameter space where they exist, these are defined by the basic reproductive number $R_0$. Both fixed points are found to be stable within their respective regions according to the linear stability analysis. The important outcome of this analysis is that the critical value $\beta_c$ (\ref{betac}) exists such that the disease-free fixed point exist only if a contact rate $\beta<\beta_c$. This indicates that, within the $SEIR$ model suggested in this study, the contact rate plays a crucial role in the possibility of the system to achieve a disease-free stationary state.

\section{Numeric solution and its fits at various stages of an outbreak}\label {III}

Stationary states of the epidemiology model describe its asymptotic behaviour at large times, and the possibility of reaching a DF state is of most importance here. However, the time-resolved dynamics of the disease dissemination is even of higher importance, as it is directly related to the lives losses and to the load put on medical service. In particular, monitoring the initial phase of the dissemination \cite{Pongkitivanichkul2020,Gotz2020,Zhang2020,Linka2020,Peirlinck2020,Kucharski2020} helps to decide on a required measures to bring its dissemination down \cite{Chatterjee2020,Nazarimehr2020,Davies2020,Bai2020,Carletti2020}.

The $SEIRS$ model, defined via the set of equations (\ref{dSdt})-(\ref{dRdt}), has no exact analytic solution for its dynamics, but we will attempt to find a suitable approximate one. We will write the equation for the total fraction of infected individuals $I'=E+I+R$ in the following form
\begin{equation}
\dot{I'} = \beta(I'-R)(1-I') - \gamma I'\label{dIpdt}
\end{equation}
One can see that at $R=0$ this equation reduces to the one for the $SI'S$ model which has an exact solution. If $R\neq 0$, the Eq.~(\ref{dIpdt}) may be solvable if the fraction of isolated individuals $R$ is some simple known function of $I'$.

To look for suitable choices for such function, we performed numerical integration of Eqs.~(\ref{dSdt})-(\ref{dRdt}). It is done via the second-order integrator
\begin{equation}
X(t+\Delta t)=X(t) + \dot{X}(t)\Delta t + \frac{1}{2}\ddot{X}(t)\Delta t^2\label{integr}
\end{equation}
for each fraction $X=\{S,E,I,R\}$, which is applied iteratively starting form the initial state, $S(0)$, $E(0)$, $I(0)$ and $R(0)$, with the time step $\Delta t$. The latter is chosen equal to one day. The equations are coupled, as far as both the first derivatives $\dot{X}$, given by Eqs.~(\ref{dSdt})-(\ref{dRdt}), and the second derivatives
%
\begin{eqnarray}
\ddot{S} & = & -\beta(\dot{E}+\dot{I})S - \beta(E+I)\dot{S} - \gamma \dot{S} \label{ddSdt}\\
\ddot{E} & = & \beta(\dot{E}+\dot{I})S + \beta(E+I)\dot{S} - (\gamma + \alpha) \dot{E} \label{ddEdt}\\
\ddot{I} & = & \alpha \dot{E} - (\gamma + \delta) \dot{I} \label{ddIdt}\\
\ddot{R} & = & \delta \dot{I} - \gamma \dot{R} \label{ddRdt}
\end{eqnarray}
at time instance $t$ depend on all variables $S$, $E$, $I$ and $R$ at the same instance $t$. It has been checked that the third-order integrator does not provide any evident improvement in accuracy as compared to the second-order one. To simplify analysis of numeric solution, we exploit the fact that $\alpha$ and $\delta$ enter expression for $R_0$ (\ref{Rnumb}) in a symmetric way and, therefore, we consider the same simplified case of $\alpha=\delta$ as above. The numeric integration can, obviously, be perforemd at any values of $\alpha$ and $\delta$.

\begin{figure}[!ht]
\begin{center}
\includegraphics[clip,width=5cm,angle=270]{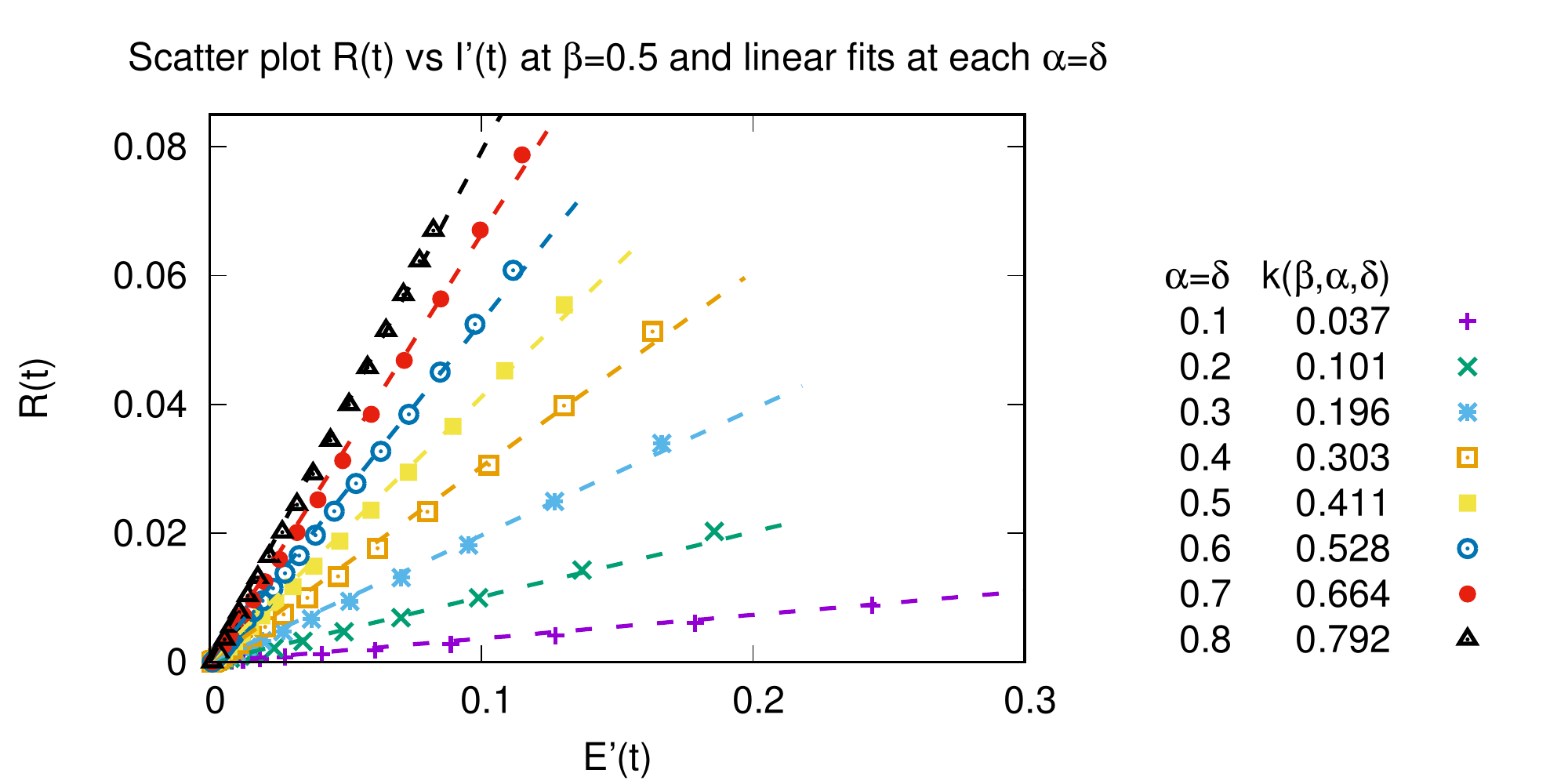}
\caption{\label{R_vs_Ip}Scatter plot $R$ vs $I'$ obtained from a numerical solution of the set of equations (\ref{dSdt})-(\ref{dRdt}) at an early-time stage. The contact rate is $\beta=0.5$ and a wide range of $\alpha=\delta$ rates is covered.}
\end{center}
\end{figure}
Let us consider the early-stage dissemination of the disease from the almost healthy initial state, $S(0)=0.999$, $E(0)=0.001$ and $I(0)=R(0)=0$, first. The total number of initially infected individuals, brought into the system from somewhere outside, is $I'(0)=0.001$ or $0.1\%$ of total populations then. The numeric integration is performed at various rates $\beta$ and $\alpha=\gamma$. To look for a suitable function that relates $R$ to $E'$, the time evolution of both fractions was presented as the scatter plots, see Fig.~\ref{R_vs_Ip}. As is seen in the plot, the linear function
\begin{equation}
R(t) \approx k(\beta,\alpha,\delta) I'(t)\label{R_vs_I_plot}
\end{equation}
provides a very good fit at early times ($t<30$) for the case of $\beta=0.5$ and a wide range of $\alpha=\delta$ rates. Fig.~\ref{R_vs_Ip} contains also the fitting results for the coefficient $k(\beta,\alpha,\delta)$ at various $\alpha=\delta$. Similar findings obtained at other values of $\beta$ lead to the approximate fitting function for $k(\beta,\alpha,\delta)$ of the following form
\begin{equation}
k(\beta,\alpha,\delta) \approx \frac{[(\alpha+\delta)/2]^{\frac32}}{2\beta}. \label{coeff_k}
\end{equation}
Its accuracy is found to be reasonable good in a wide interval from $\beta=0.1$ to $0.5$ and $\alpha=\delta$ ranging from $0$ to $1$. The cases $\beta=0.1$ and $0.5$ are shown in Fig.~\ref{k_fit}, where the approximate expression (\ref{coeff_k}) is plotted via dashed lines and the results for $k(\beta,\alpha,\delta)$ obtained by numeric integration -- via squares.  

\begin{figure}[!ht]
\begin{center}
\includegraphics[clip,width=5cm,angle=270]{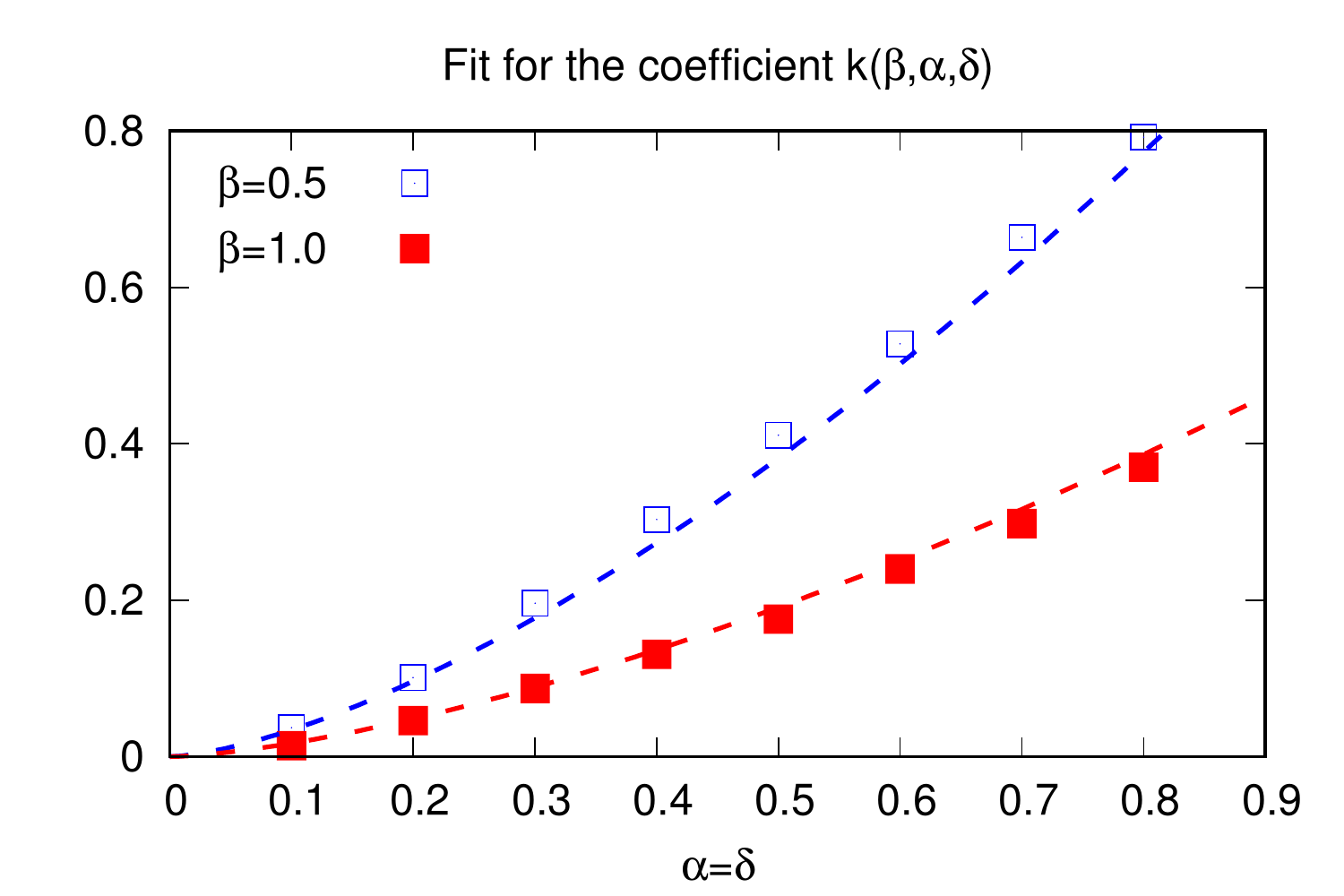}
\caption{\label{k_fit}Check for the accuracy of the approximate expression (\ref{coeff_k}) for $k(\beta,\alpha,\delta)$ at a range of $\beta$ and $\alpha=\delta$ indicated in the plot.}
\end{center}
\end{figure}
The approximate expression for $R(t)$ given by Eqs.~(\ref{R_vs_I_plot}) and (\ref{coeff_k}), can be substituted into Eq.~(\ref{dIpdt}) now resulting in the equation
\begin{equation}
\dot{I'} = \beta'I'(1-I') - \gamma I' \label{dIpdt2}
\end{equation}
where $\beta'=\beta(1-k(\beta,\alpha,\delta))$. One recognizes now the equation for the $SI'S$ model with the scaled contact rate $\beta'$. It is lower than $\beta$ as far as part of infected individuals are isolated in the group $R$ and do not infect susceptible individuals via social contacting. The solution of this equation is well known
\begin{equation}
I'(t) = \frac{I'(0)(1-\gamma/\beta')}{I'(0) + (1-\gamma/\beta'-I'(0))\exp[-\beta'(1-\gamma/\beta')t)] } \label{Ip_solut}
\end{equation}
\begin{figure}[!ht]
\begin{center}
\includegraphics[clip,width=5cm,angle=270]{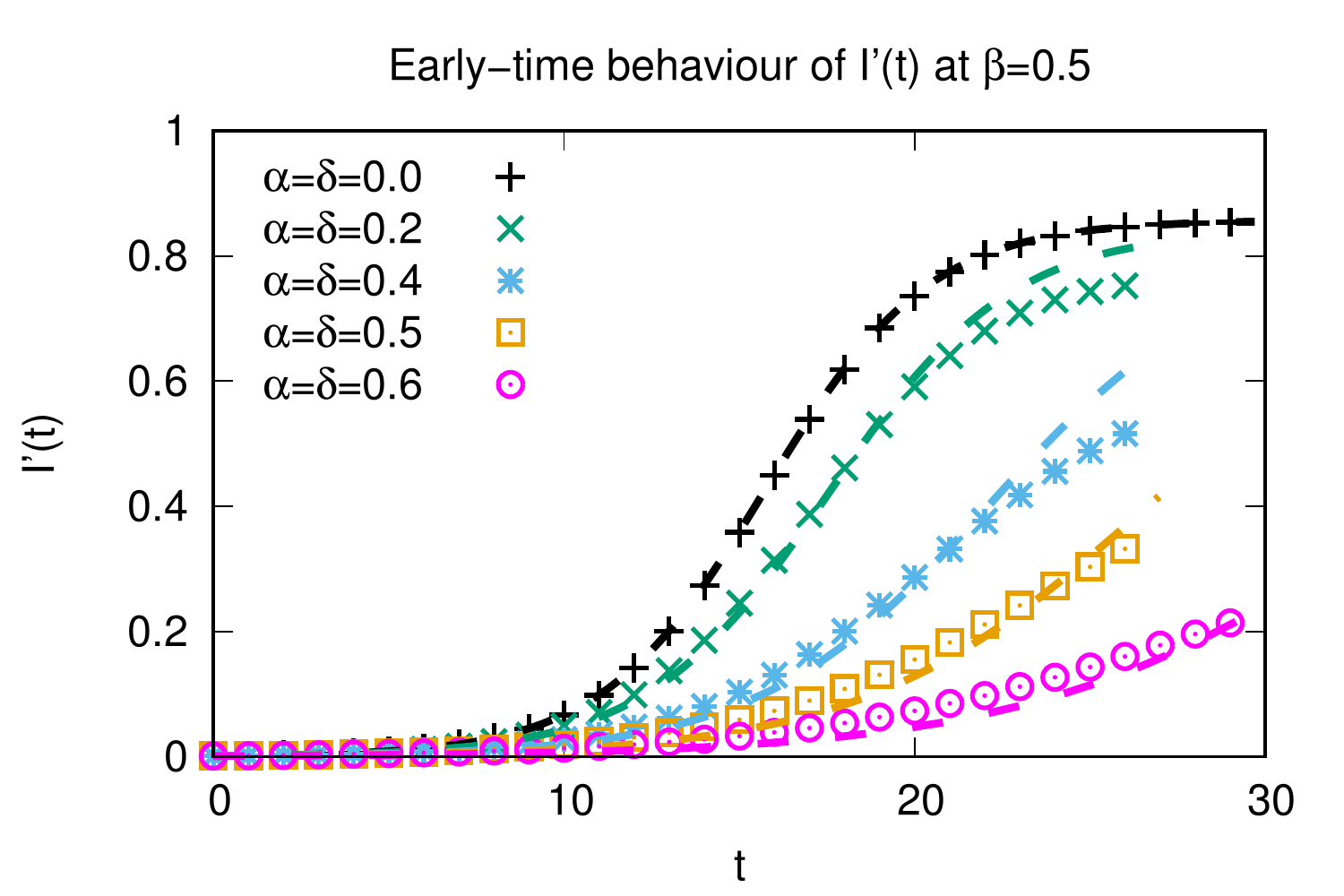}
\caption{\label{Ip_evol_increase} Data points: numeric result for the time evolution of the total fraction of infected individuals $I'(t)$ starting from an initial state with $I'(0)=0.001$ at $\beta=0.5$ and a range of $\alpha=\delta$. Lines: approximate solution given by Eq.~(\ref{Ip_solut}).}
\end{center}
\end{figure}
The approximate solution (\ref{Ip_solut}) is found to reproduce the early-time dynamics of $I'(t)$ from the almost healthy state ($I'(0)=0.001$) rather well and in a wide range of $\beta$ and $\alpha=\delta$. This is demonstrated in Fig.~\ref{Ip_evol_increase} for the case of $\beta=0.5$ and $\alpha=\delta$ that range from $0$ (where the approximate solution is exact) to $0.6$.

\begin{figure}[!ht]
\begin{center}
\includegraphics[clip,width=5cm,angle=270]{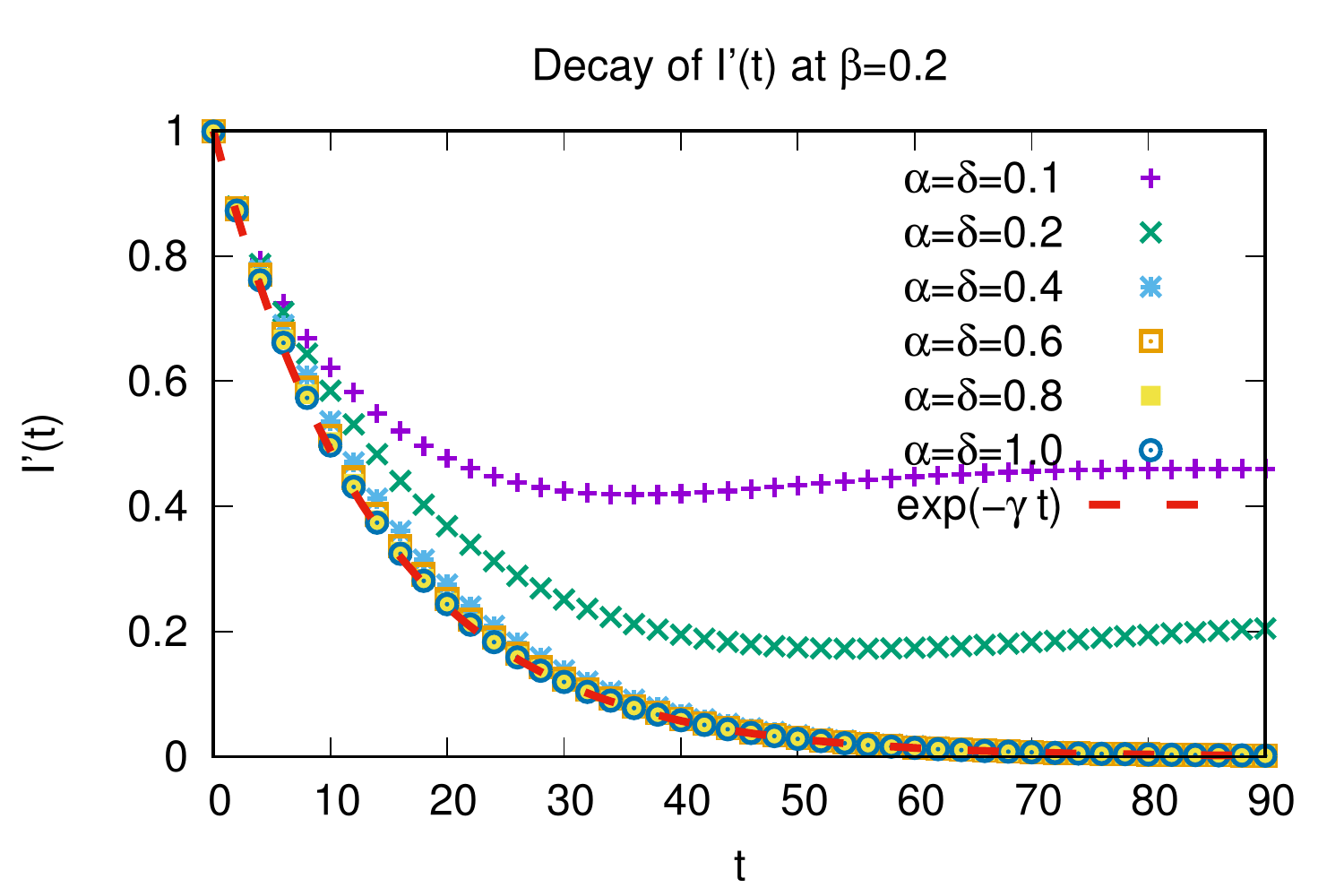}
\caption{\label{Ip_evol_decay} Data points: numeric result for the time evolution of the total fraction of infected individuals $I'(t)$ starting from an initial state with $I'(0)=0.999$ at $\beta=0.2$ and a range of $\alpha=\delta$. Lines: approximate solution given by Eq.~(\ref{Ip_gamma_solut}).}
\end{center}
\end{figure}
Let us now consider the case when the disease decays from the all-infected state with $S(0) = 0.001$ and $E(0)=I(0)=R(0)=0.333$ because of a suitable choice of the model parameters (e.g. as the result of the quarantine measures, massive testing and efficient isolation of infected individuals). This, of course, is possible only if the set of $\beta$ and $\alpha=\delta$leads to the reproductive number $R_0\leq 1$, where the latter is given by Eq.~(\ref{Rnumb}). Using fit to the numeric integration data, we found that the decay dynamics of $I'(t)$ can be approximated well by the exponential function
\begin{equation}
I'(t) = I'(0)\exp(-\gamma t) \label{Ip_gamma_solut}
\end{equation}
as demonstrated in Fig.~\ref{Ip_evol_decay} for the case of $\beta=0.2$ and $\alpha=\delta$ ranging from $0.1$ to $1$. At this value for $\beta$, $R_0$ crosses $1$ at $\alpha=\delta\approx 0.29$ and the exponential decay (\ref{Ip_gamma_solut}) is very accurate, indeed, at $\alpha=\delta \geq 0.29$ (see respective plots in Fig.~\ref{Ip_evol_decay}). The expression (\ref{Ip_gamma_solut}) is the solution of the equation
\begin{equation}
\dot{I'} = - \gamma I' \label{dIpdt3}
\end{equation}
which is obtained from Eq.~(\ref{dIpdt}) at $\beta=0$. This indicates that the $\gamma I'$ term prevails over the $\beta(I'-R)(1-I')$ one for the decay dynamics of the $SEIRS$ model considered here. Therefore, if the reproductive number $R_0\leq 1$, then the disease can be brought down more quickly only by the increase of a curing rate $\gamma$ (by finding effective medication), but not by changing the other parameters related to the contact, identification and isolation rates of infected individuals.

\begin{figure}[!ht]
\begin{center}
\includegraphics[clip,width=5.5cm,angle=270]{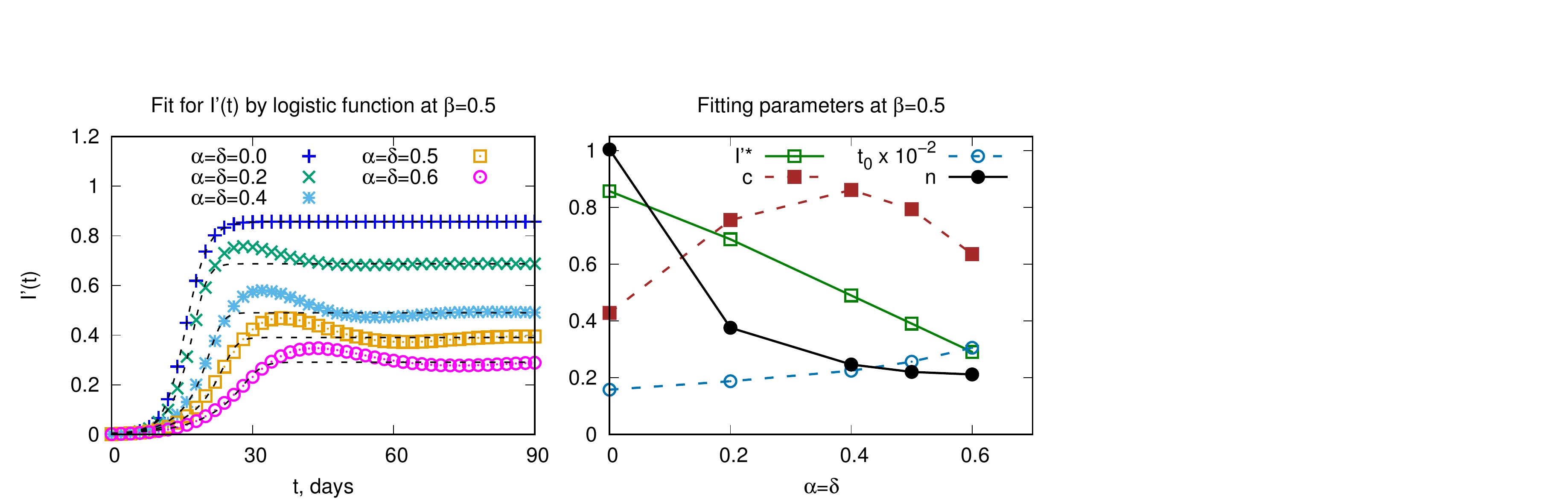}
\caption{\label{RG_logistic_fit} Left frame: Fit of the $I'(t)$ evolution via modified logistic function (\ref{RG_logistic_form}) at $\beta=0.5$ and a range of $\alpha=\delta$. Right frame: dependence of the set of fitting parameters on $\alpha=\delta$.}
\end{center}
\end{figure}
As an attempt to fit the evolution of $I'(t)$ both at the early times of an outbreak and at it saturation in an endemic state, we used the modified logistic function
\begin{equation}
I'(t) = {I'}^*\left[1+\exp(-c(t-t_0))\right]^{-n} \label{RG_logistic_form}
\end{equation}
suggested by Pongkitivanichkul et al. \cite{Pongkitivanichkul2020} as the result of their analysis of the epidemic data reported by the World Health Organization, and motivated by the renormalization group framework. Here ${I'}^*=1-S^*$, where $S^*$ is given by Eq.~(\ref{Sstat}), $c$ and $t_0$ define a time scale for the $I'(t)$ evolution, whereas $n$ is termed in Ref.~\cite{Pongkitivanichkul2020} as an asymmetry in the modified logistic function, which determines the characteristic of the epidemic at an early stage and is found either equal or less than one. The fits, performed for the numerical solution of the $SEIRS$ model considered here, are shown at $\beta=0.5$ and a range of $\alpha=\delta$ from $0$ to $0.6$ as black dashed lines in the left frame of Fig.~\ref{RG_logistic_fit}. The analytic form (\ref{RG_logistic_form}) is found to work well at all times, except straightening up the bumps before the plateau is reached, at $\alpha=\delta>0$. The dependence of the fitting parameters ${I'}^*$, $c$, $t_0$ and $n$ on $\alpha=\delta$ are shown in the right frame of the same plot. Leaving out the amplitude ${I'}^*$, discussed above, let us note that the value of $c$ is found inside the interval from $0.4$ to $0.8$ without obvious systematic trend, whereas the characteristic time $t_0$ increases from about $19$ to $30$ days when $\alpha=\beta$ raised from $0$ to $0.6$. The strongest dependence is found for the asymmetry $n$, which decreases from $1$ down to $0.25$ and seems to saturate at larger $\alpha=\beta$. This result is revelant in the context of the role of this parameter for description of the maturing and the growth-dominated phase of the pandemy, as discussed in Ref.~\cite{Pongkitivanichkul2020}. Deeper analysis is, however, beyond this report and will the subject of the following studies. 

To conclude this section, we performed numerical integration of the $SEIRS$ model here and found that, at early stage of an outbreak, the fraction $R$ of isolated infective individuals is linearly proportional to the total fraction $I'=E+I+R$ of infected individuals. This enabled us to map the $SEIRS$ model onto the $SI'S$ one with effective contact rate $\beta'<\beta$ and obtain the approximate analytic solution for the $SEIS$ model valid at the early time of the outbreak. The decay dynamics is found to be dominated by the recovery rate $\gamma$, which is fixed in this model. The evolution of $I'(t)$ in the whole time interval can be fitted well with the modified logistic function of Pongkitivanichkul et al. \cite{Pongkitivanichkul2020} leading to the possibility of further analysis of the outbreak merits.

\section{The effects of quarantine measures and of their relaxation}\label {IV}

In the previous section we considered both the early-time dissemination and the decay of the disease as described by the $SEIRS$ model that reflects characteristic features of the COVID-19 virus. Numeric solution (\ref{integr}) is a quick and efficient way to do such analysis for any particular choice of the model parameters $\beta$, $\alpha$, $\delta$ and $\gamma$. On a top of this, we also obtained approximate analytic solutions (\ref{Ip_solut}) and (\ref{Ip_gamma_solut}) at $\alpha=\delta$ for the early-time dissemination, for the decay regimes and the fit to the modified logistic function (\ref{RG_logistic_form}). This analysis is performed assuming that the evolution of the system occurs at fixed values for all model  parameters $\beta$, $\alpha$, $\delta$ and $\gamma$. The real-life situation is rather different, as the governments and societies react dynamically on the virus dissemination by undertaking appropriate measures. The latter can be modelled via the dynamic changes of  the model parameters in a course of the disease dissemination.

The main problem faced by all countries during the initial stages of the COVID-19 pandemic was to slow it down, which will reduce the strain on the medical care system. This was tackled by means of a number of quarantine measures, both locally (closure of sporting and cultural mass attendance events, hotel service, social distancing, etc.) and globally (cancellation of international flight and train services, closure of tourism, etc.).\cite{Linka2020,Mushayabasa2020,Linka2020,Davies2020,Choi2020,Contreras2020,Peirlinck2020,Gatto2020,Zhang2020,Zhang2020,Choi2020}.

 However, at the subsequent stages of the pandemic, it became evident that the quarantine measures have a number of serious economic implications. Therefore, the focus shifted gradually to addressing the questions when and to which extend the quarantine measures can be relaxed to revive economy but, simultaneously, to keep the dissemination of the disease under control \cite{Chowdhury2020,Block2020,Rawson2020,Lopez2020}. The effects of a quarantine and of it relaxation will be discussed for the case of the $SEIRS$ model in this section. We must disclaim and remind again that this model reproduces the extreme case when no immunity can be acquired against the disease, and that we do not aim on reproducing any particular case (country, town) or provide practical aids with numbers. Instead, we examine the general effects and trends related to quarantine depending on the model parameters set.

We start from discussing the effect of introducing permanent quarantine measures on the dynamics of the disease dissemination. The following algorithm is used. We start from the state with a small fraction of unidentified infected individuals $E_0$ (supposedly brought to the unaware community from outside), whereas forth $I$ and $E$ are initially equal to zero. At early times the dissemination of disease is unrestricted and is characterised by a ``normal'' contact rate $\beta$, being characteristic for a given community. When, at a certain time instance, the fraction of newly identified individuals per day, $\dot{I}$, reaches the quarantine threshold value of $\Delta$, the community is switched into a quarantine mode characterised by a lower contact rate $\beta_2<\beta$.  

For reading convenience, we display all fractions in $\%$ of the total population. We present the results for three separate cases of $\alpha=\delta=0.05$, $0.1$ and $0.2$, which model poor, intermediate and high identification and isolation rates, respectively. All other parameters, $E_0=0.001\%$, $\beta=0.25$, $\beta_2=0.05-0.25$ and $\Delta=0.01\%$, are the same in all cases. We also remark that the findings are similar at other values of $\beta$ if the other rates are scaled accordingly. For the sake of brevity, we display the dynamics of the isolated infected fraction $R$ only, which is important as an estimate of a strain put on the medical care system. The unidentified infected fraction, $E$, cannot be evaluated in a real life, whereas the identified infected fraction, $I$, can be interpreted as the ``transit'' state for the individuals that are identified as infected but not isolated yet and, hence, of a lesser importance. In any case, the behaviour of both $E$ and $I$ is found to follow closely that of $R$. 

\begin{figure}[!ht]
\begin{center}
\includegraphics[clip,width=5.0cm,angle=270]{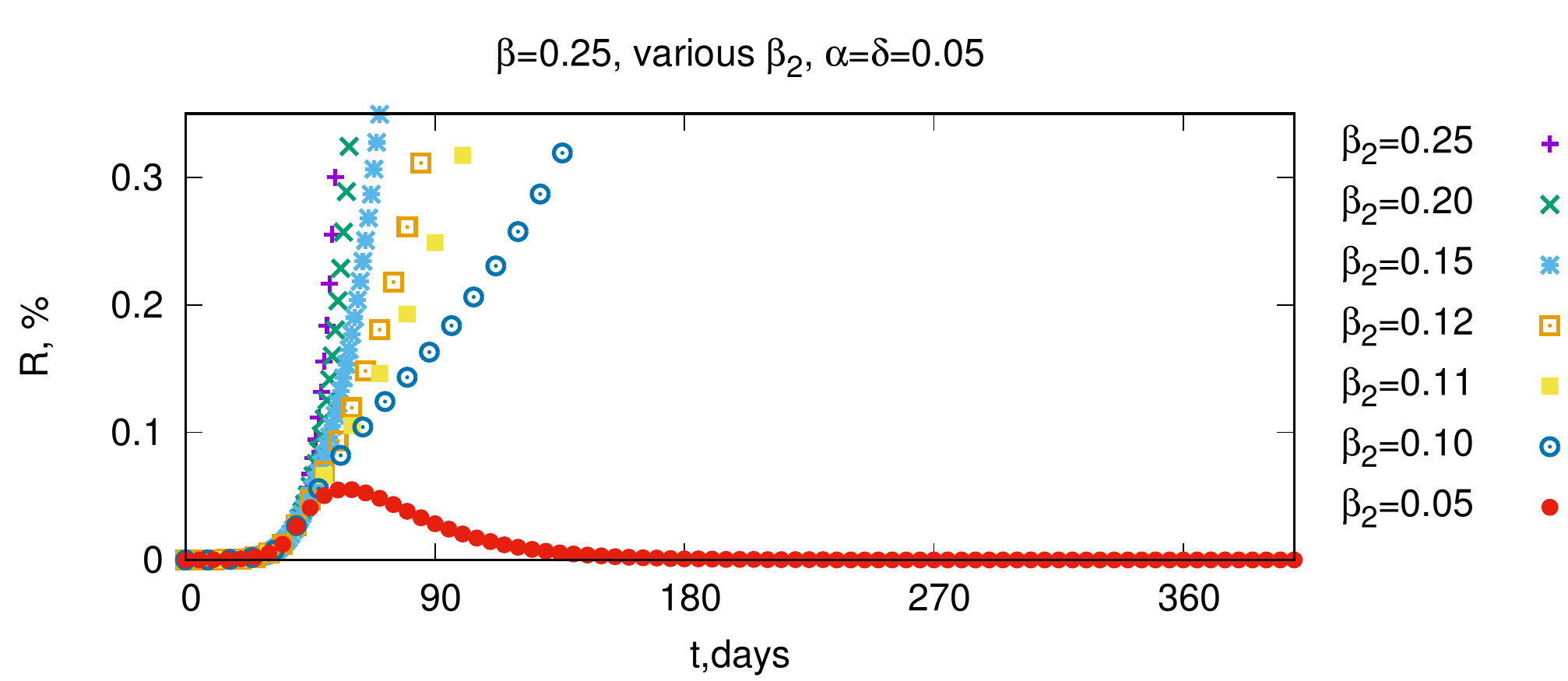}
\includegraphics[clip,width=5.0cm,angle=270]{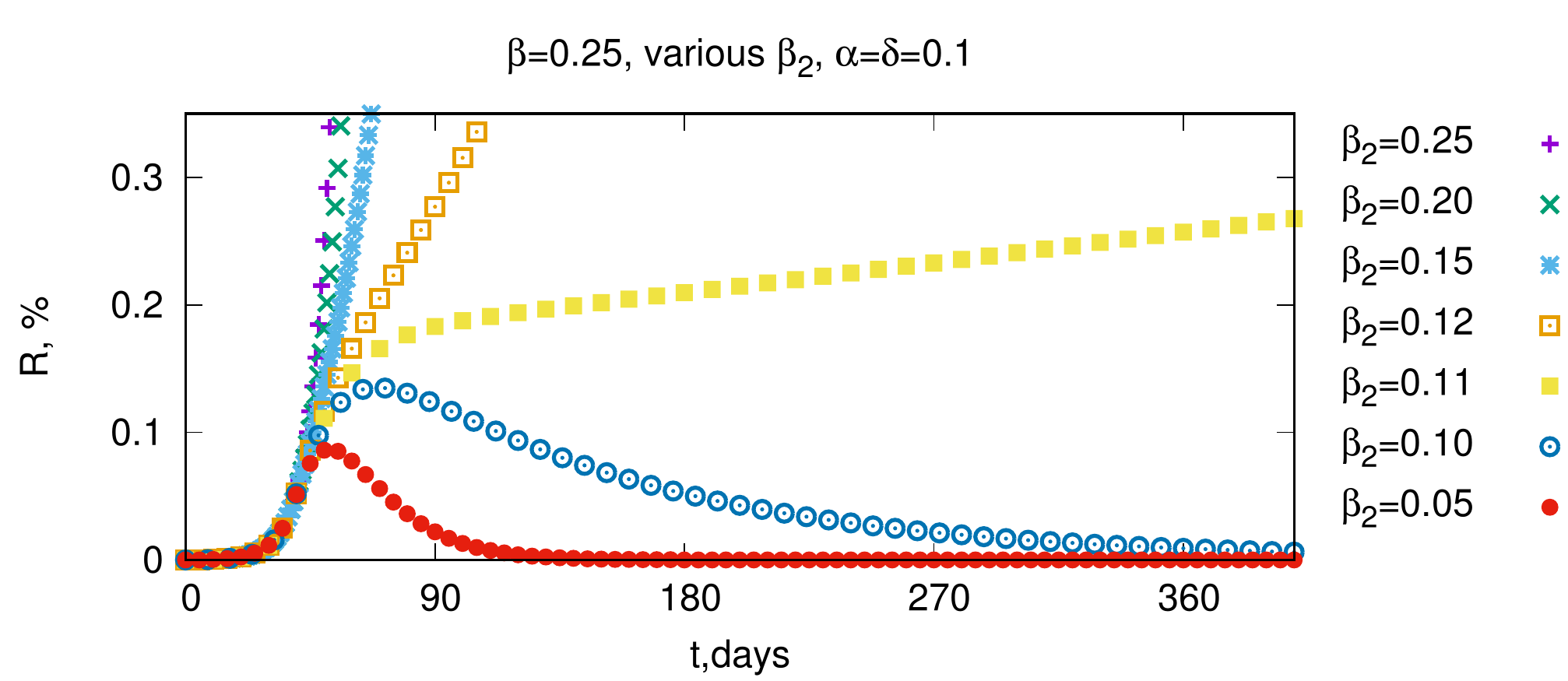}
\includegraphics[clip,width=5.0cm,angle=270]{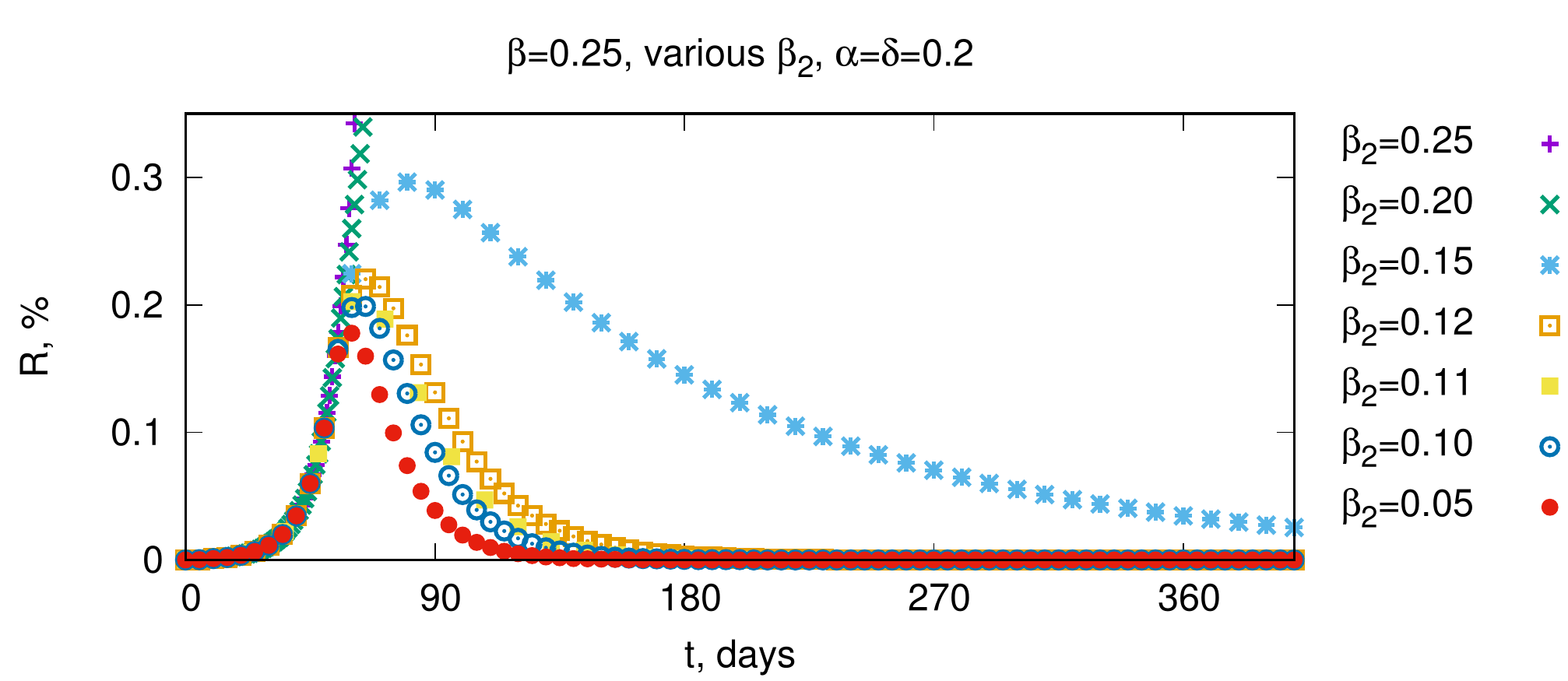}
\caption{\label{quarant_on}Top frame: time evolution of $R(t)$, as affected by introducing the quarantine measures of various severity ($\beta_2$ ranges from $0.05$ up to $5$), at very low identification and isolation rates $\alpha=\delta=0.05$. Middle and bottom frames: the same for higher values of $\alpha=\delta=0.1$ and $\alpha=\delta$, respectively.}
\end{center}
\end{figure}
The top plot in Fig.~\ref{quarant_on} shows the case of a low identification and isolation rates, $\alpha=\delta=0.05$ and various quarantine levels, from a very strict one, $\beta_2=0.05=\beta/5$, to a no quarantine being introduced, $\beta_2=0.25=\beta$. As one can see in this plot, the decay of the $R$  fraction to zero at long times is possible only at a very strict quarantine measures, $\beta_2=0.05$. Twofold increase of the identification and isolation rates to $\alpha=\delta=0.1$ (see, middle plot in Fig.~\ref{quarant_on}), enables to achieve this at less strict quarantine measures characterised by a higher contact rate $\beta_2=0.1$. Yet another twofold increase of $\alpha=\delta$ up to $0.2$ (shown in a bottom plot in Fig.~\ref{quarant_on}) increases threshold value of $\beta_2$ further, up to $0.15$. This demonstrates that the strictness of quarantine measures in the $SEIRS$ model is inversely proportional to the identification and isolation rates. Let us also note that at $\beta_2<\beta_{2,\mathrm{th}}$, the dynamics of the $R$ changes very weakly upon further decrease of $\beta_2$. This indicates that imposing quarantine measures stronger than these characterised by$\beta_{2,\mathrm{th}}$ has no point, as it only increases associated financial burden with no real benefit of better control the pandemy. 

\begin{figure}[!ht]
\begin{center}
\includegraphics[clip,width=6.0cm,angle=270]{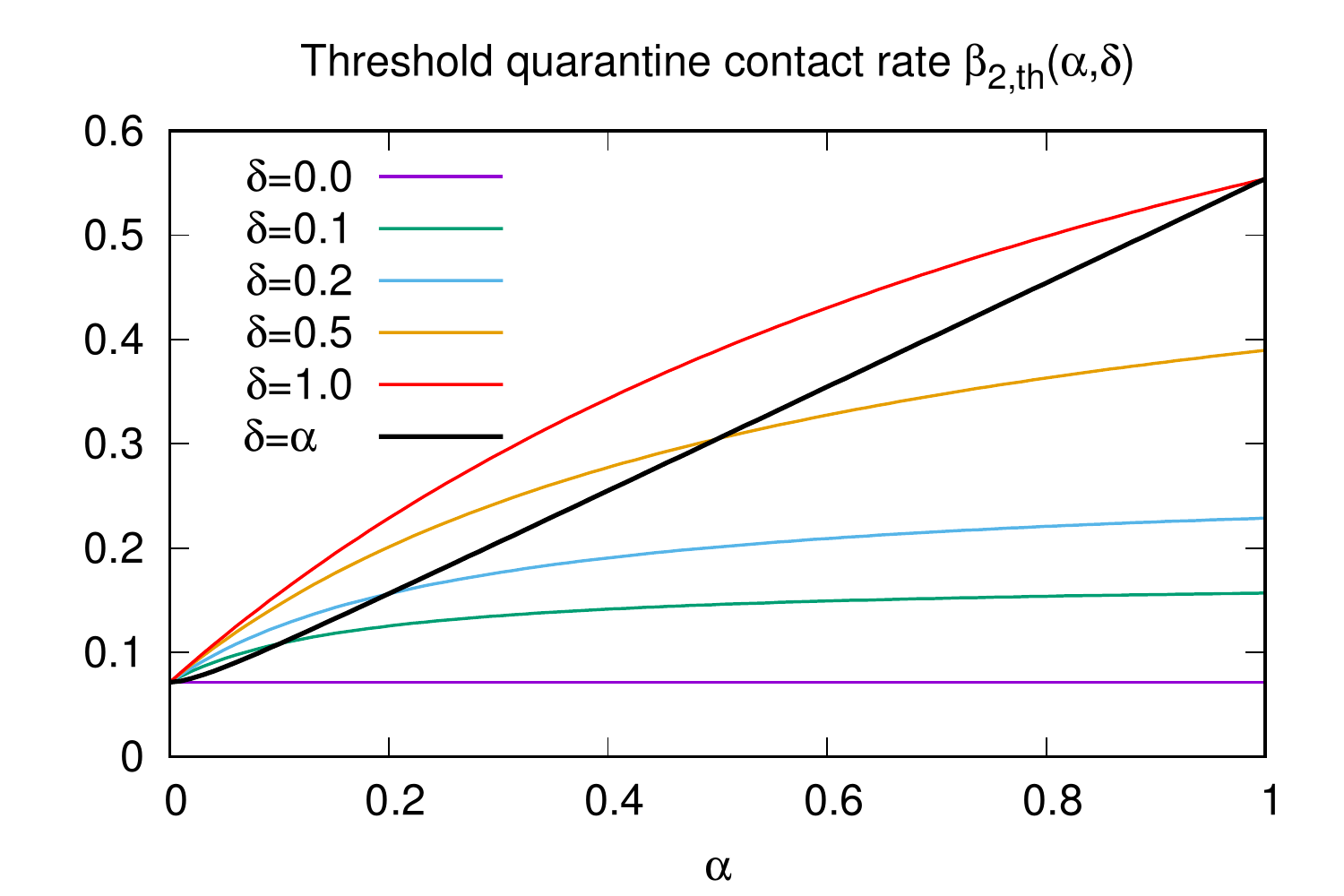}
\caption{\label{beta2_th}The threshold quarantine contact rate, $\beta_{2,\mathrm{th}}$, as a function of $\alpha$ and $\delta$.}
\end{center}
\end{figure}
Exact expression for the threshold value, $\beta_{2,\mathrm{th}}$, at which $R$ has a zero stationary state, can be obtained from the equation $R_0=1$, where $R_0$ is given by Eq.~(\ref{Rnumb})
\begin{equation} 
\beta_{2,\mathrm{th}}(\alpha,\delta)=\frac{(\gamma+\alpha)(\gamma+\delta)}{\gamma+\alpha+\delta}.\label{beta2_expr}
\end{equation} 
Given that this expression is symmetric with respect to $\alpha$ and $\delta$, we analyse $\beta_{2,\mathrm{th}}(\alpha,\delta)$ as a function of one of its arguments while another is fixed plus the special case of $\alpha=\delta$ which is exploited throughout this study. This is done in Fig.~\ref{beta2_th}, and we see that the decrease of one of its arguments (in this case this is $\delta$) below $0.2$ decreases $\beta_{2,\mathrm{th}}(\alpha,\delta)$essentially regardless of the value of the other argument. With respect to this, the special case of $\alpha=\delta$ turns to be quite optimal, as: (i) the curve for $\beta_{2,\mathrm{th}}(\alpha,\delta=\alpha)$ is relatively high approaching the case when one of the arguments equals to $1$ and (ii) the value of both arguments can be kept balanced without much demand on either identification or isolation rate to stay very high. In the special case $\alpha=\delta$, the dependence of $\beta_{2,\mathrm{th}}(\alpha,\delta=\alpha)$ on $\alpha$ is close to linear. Indeed, the expansion of Eq.~(\ref{beta2_expr}) at small $\gamma$ yields linear dependence in $\alpha$
\begin{equation} 
\beta_{2,\mathrm{th}}(\alpha,\delta)\approx\frac{3}{4}\gamma + \frac{\alpha}{2}.\label{beta2_appr}
\end{equation} 
This expression provides simple means of evaluation of minimum required quarantine measures depending on the current identification and isolation rates.

We will switch now to the issues when and how the quarantine measures can be relaxed while keeping the disease dissemination under control, where we follow closely the ideas from Ref.~\cite{Rawson2020}. In particular, the quarantine measures (contact rate is reduced to $\beta_2$) are introduced if the fraction of newly identified infected individuals per day are $\dot{I}\geq\Delta$. However, if, as the result of quarantine measures, the disease dissemination decays and $\dot{I}\leq\epsilon$, where $\epsilon$ can be termed as the quarantine relaxation threshold, then the quarantine is relaxed. Relaxation means reversion of the contact rate back to $\beta$, a ``normal life'' one.

\begin{figure}[!ht]
\begin{center}
\includegraphics[clip,width=5.0cm,angle=270]{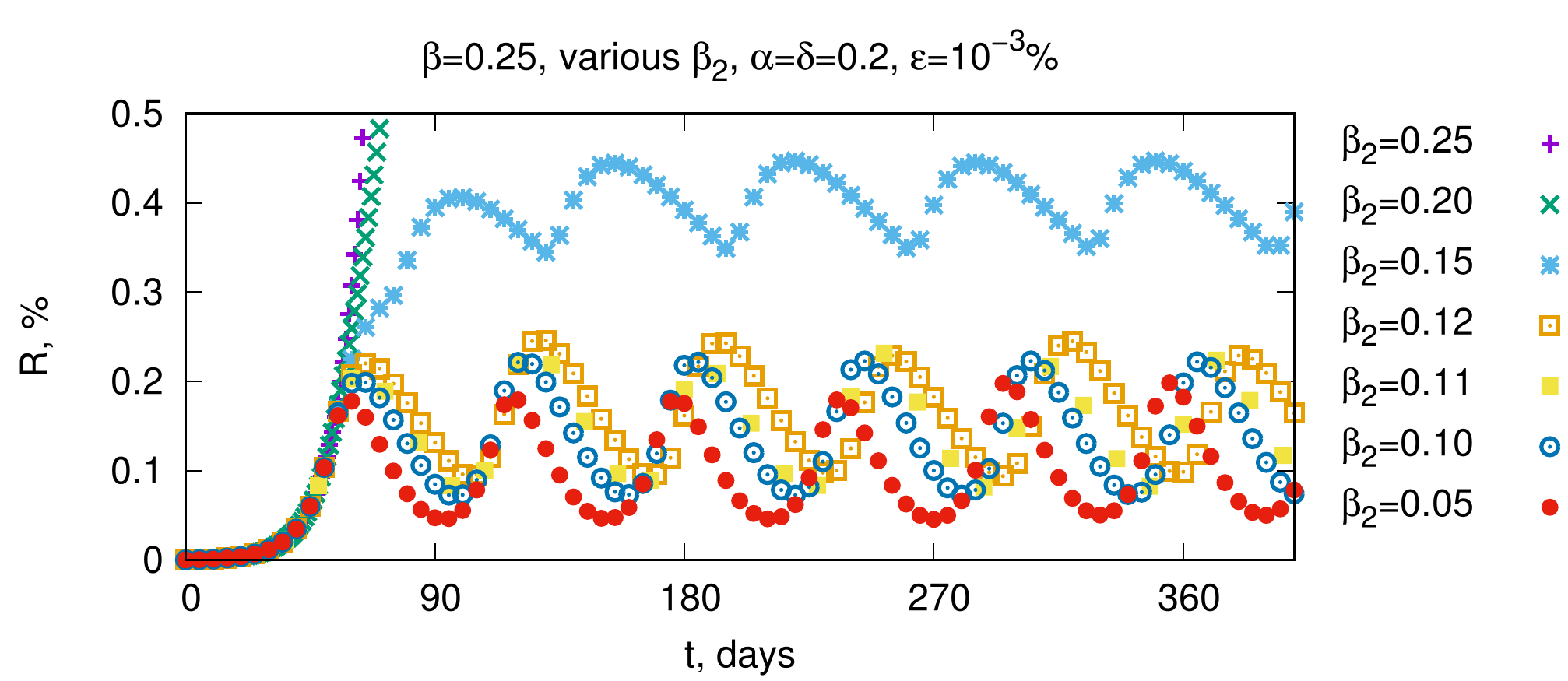}
\includegraphics[clip,width=5.0cm,angle=270]{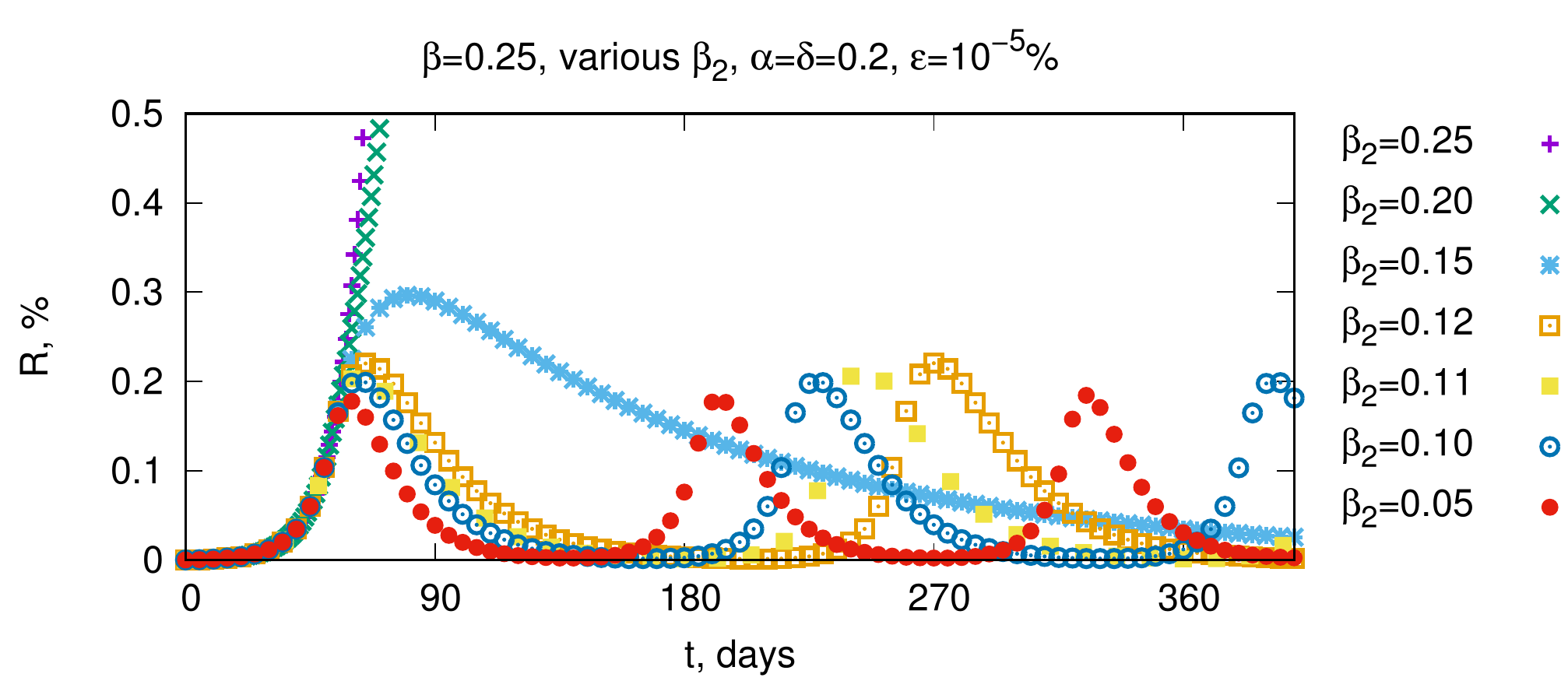}
\includegraphics[clip,width=5.0cm,angle=270]{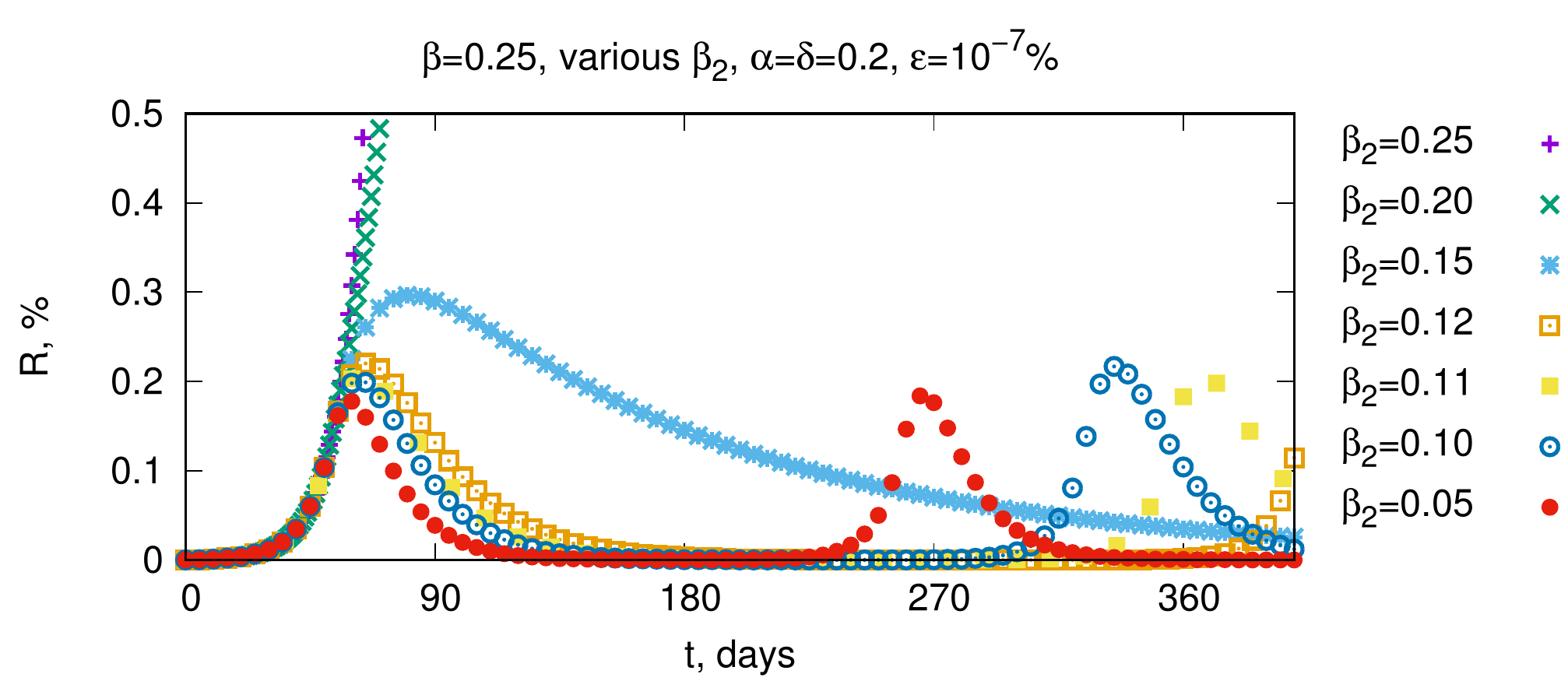}
\caption{\label{quarant_relax}Top frame: time evolution of $R(t)$ in the the periodic ``quarantine on'' -- ``quarantine off'' scenario at fixed $\alpha=\delta=0.2$ and relatively large threshold value $\epsilon$. Middle and bottom frames: the same at more moderate and low threshold value $\epsilon$, as indicated in the figures.}
\end{center}
\end{figure}
We concentrate here on the resulting dynamics for the $R$ fraction depending on the choice for $\epsilon$ for the same set of parameters: $\beta=0.25$, $\alpha=\delta=0.2$ and a range of $\beta_2$ as shown previously in the bottom frame of Fig.~\ref{quarant_on}. The results are shown in Fig.~\ref{quarant_relax}. Top plot in this figure represents relatively early relaxation of quarantine measures, $\epsilon=10^{-3}\%$. This results in a wave-like oscillations for $R$ with no time intervals where $R$ drops essentially below $0.1\%$. With the decrease of  $\epsilon$ down to $10^{-5}\%$, such intervals appear and are of approximate duration of $50-60$ days (see, middle plot in the same figure). With the further decrease of $\epsilon$ down to $10^{-7}\%$, their duration increase about twice, to about $100$ days (see, bottom plot there). Therefore, lowering the quarantine relaxation threshold $\epsilon$ about two orders of magnitude increases the length of the almost disease-free time intervals about twice. During these intervals, the economics works normally and can be essentially revived. Let us also remark that the threshold value for quarantine contact rate $\beta_{2,\mathrm{th}}=0.156$ for the given choice of $\alpha=\delta=0.2$. And, as follows from the middle and bottom all plots in Fig.~\ref{quarant_relax}, duration of the almost disease-free intervals is maximal when $\beta_2\approx 0.8\beta_{2,\mathrm{th}}$ and decays when $\beta_2$ decreases. This indicates, that, at least within this model, an optimal quarantine contact rate $\beta_2$ exists such that it leads to the longest disease-free interval when the quarantine measures are relaxed. This contact rate is lower but close to the threshold contact rate $\beta_{2,\mathrm{th}}$.

The height of the second and all the following waves for the disease dissemination in Fig.~\ref{quarant_relax} is found to be the same. This is the consequence of the fact, that the only way to bring the disease down in the $SEIRS$ model is to transfer infected individuals into the $R$ fraction, where they are isolated and do not dissemination infection further. That is why the values of the $\alpha$ and $\delta$ rates should be sufficiently high, as far as they drive this transfer. From the practical point of view, if the following waves of a pandemy appear, then one seeks the ways to bring their height down. The first way to achieve this is to take into account the immunity factor (permanent or temporal), acquired either naturally or via vaccination in a course of the disease dissemination. Immunised individuals will be resilient to the disease and will not dissemination it further. This factor is beyond the scope of this study and is a subject for a future work. Another ways of bringing the consequent waves down is to dynamically adjust quarantine threshold and/or the identification and isolation rates. These approaches are discussed below.

\begin{figure}[!ht]
\begin{center}
\includegraphics[clip,width=5.0cm,angle=270]{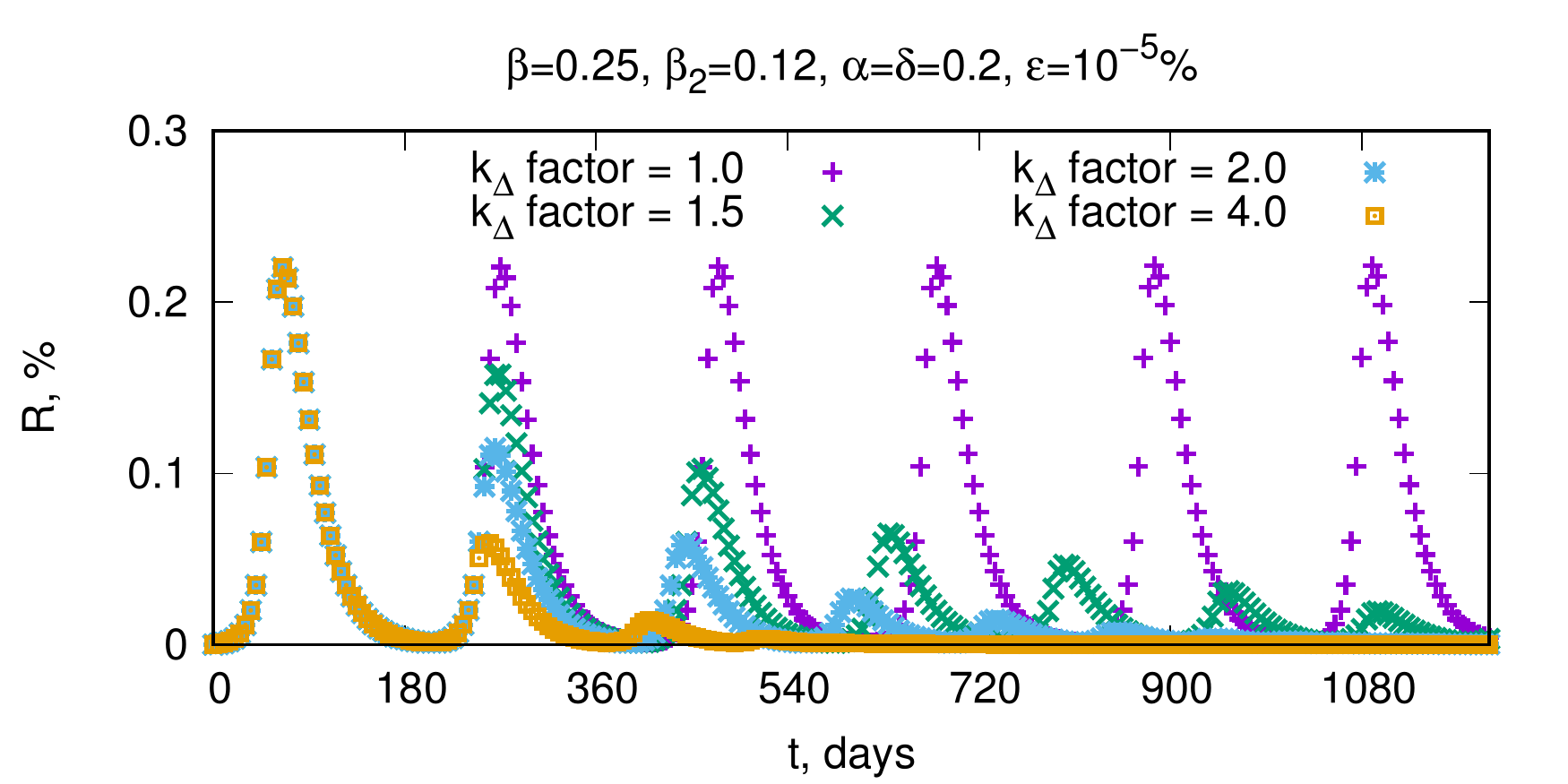}
\includegraphics[clip,width=5.0cm,angle=270]{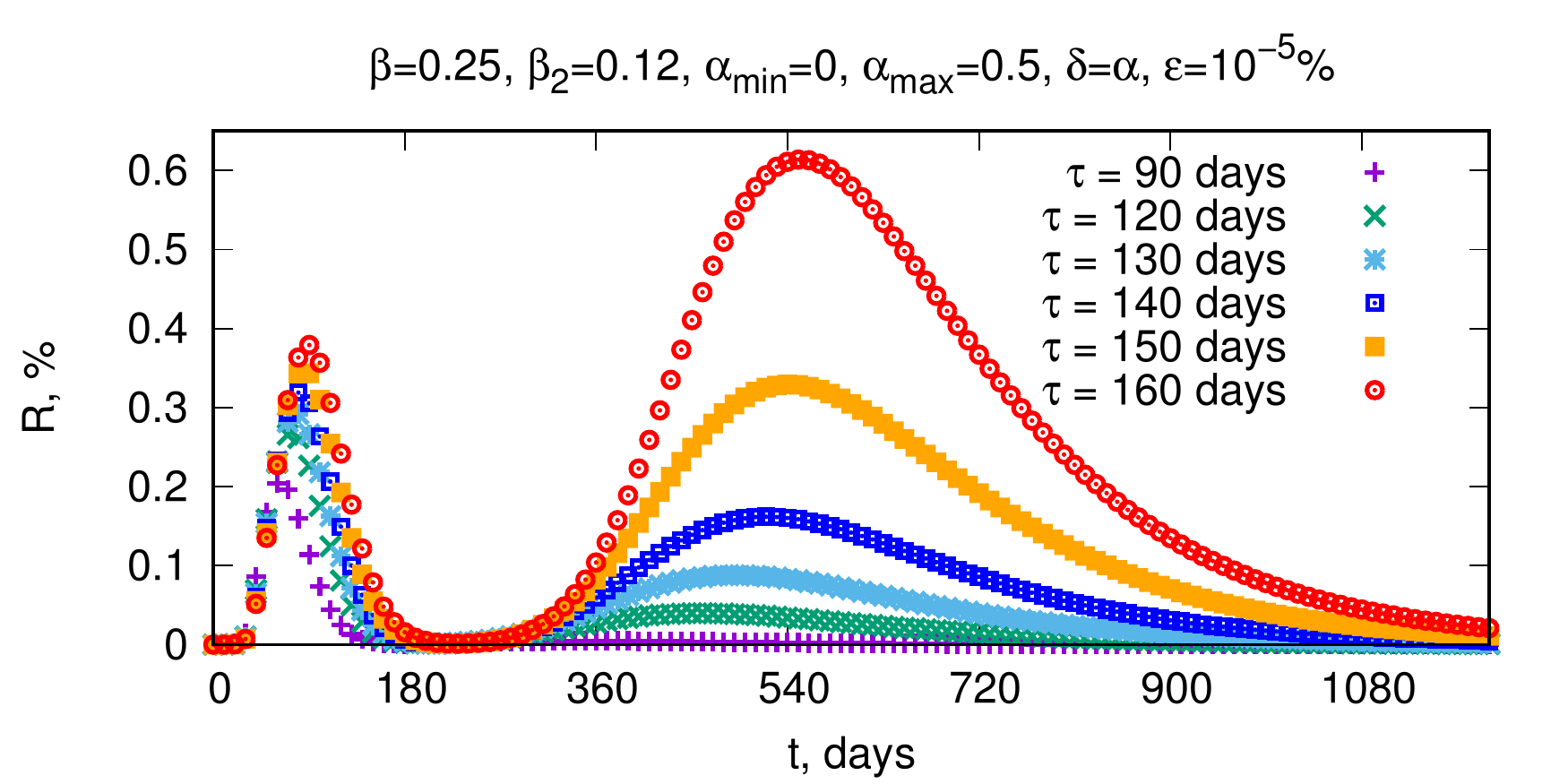}
\caption{\label{waves}Top frame: time evolution of $R(t)$ in the periodic ``quarantine on'' -- ``quarantine off'' scenario at fixed $\alpha=\delta=0.2$ using dynamic changes in $\epsilon$ according Eq.~(\ref{deltan}) at various $k_\Delta$ indicated in the figure. Bottom frame: the same but using fixed $\epsilon$ and dynamic change of $\alpha=\delta$ according to Eq.~(\ref{alphadeltat}) with the parameters provided in the figure.}
\end{center}
\end{figure}
The following procedure is used for dynamic adjustment of the quarantine threshold $\Delta$. The same basic algorithm is used as discussed with relation to Fig.~\ref{quarant_relax}. On a top of that, the quarantine threshold $\Delta_n$ for the $n$th wave obey the geometric progression
\begin{equation} 
\Delta_n=\frac{\Delta_0}{k_\Delta^n},\label{deltan}
\end{equation} 
$\Delta_1=0.01\%$, as used above, and $k_\Delta$ is the threshold factor. This means that for each subsequent wave the quarantine measures are introduced earlier than for the previous one. The result of numeric solution is shown in the top frame of Fig.~\ref{waves} at various threshold factors $k_\Delta$. One can see that the height of the second and of the following waves decrease and the effect is stronger for larger $k_\Delta$. At large times, however, $\Delta\to 0$ and the quarantine must be introduced instantly with no periods of its relaxation. Therefore, this approach may be interpreted as a bit artificial or, at least, impractical. The other way to bring the following waves down is by dynamic adjustment of the identification $\alpha$ and isolation $\delta$ rates. This might be more realistic, as it reflects the fact that with the progress of the pandemy community became more aware of it and more relevant tests are developed and available, as well as more hospital space in allocated for isolation of infected individuals. Following previous findings, especially these shown in Fig.~\ref{beta2_th}, we consider the case of $\delta=\alpha$ as the most optimal one and restrict our analysis to it. Dynamic adjustment of relevant rates are given by logistic function:
\begin{equation} 
\alpha(t)=\alpha_{\mathrm{min}} + (\alpha_{\mathrm{max}}-\alpha_{\mathrm{min}})\big(1-e^{-t/\tau}\big), \delta(t)=\alpha(t)\label{alphadeltat}
\end{equation} 
where $\alpha_{\mathrm{min}}=\alpha(0)$, $\alpha_{\mathrm{max}}=\alpha(\infty)$ and $\tau$ is the time constant that provides a timescale for the increase of $\alpha$ and $\delta$. The case of $\alpha_{\mathrm{min}}=0$ and $\alpha_{\mathrm{max}}=0.4$ is shown in the bottom of Fig.~\ref{waves} at various time constants $\tau$. As one can see, prompt reaction of a community ($\tau\leq 90$ days) completely eliminates the second and the following waves, whereas the sloppy reaction with higher $\tau$ leads to wide and high second wave, which height is comparable to (at $\tau=150$ days) or higher then (at $\tau=160$ days) the first wave. One can see that quite moderate increase of $\tau$ from $130$ to $160$ results in quite dramatic increase of the second wave height. One can conclude that the dependence of the height for the second and the following waves of the disease dissemination depends non-linearly on the model parameters related to identification and isolation of infected individuals. Consequently, slow and inefficient testing measures can be dangerous resulting in broad and high second and the following waves of the disease dissemination.

\section{Conclusions}\label {V}

We propose here the $SEIRS$ compartmental epidemiology model that is based on such features of the COVID-19 disease as: abundance of unidentified (asymptomatic or with mild symptoms) infected individuals, the absence of a vaccine, existence of a few known variants of the virus and their possible mutation, and reported cases of reinfection after recovery from the disease. Therefore, the model contains four compartments: of susceptible $S$, unidentified infected $E$, identified infected $I$ and isolated infected $R$ individuals. All types of infected individuals recover only in a natural way with the same average recovery time and with the possibility of the reinfection. Model parameters involve: the contact $\beta$, identification $\alpha$ and isolation $\delta$ rates, as well as fixed curing rate $\gamma$.

We found two stationary states (fixed points) for the set of differential equations of the model: the disease-free and the endemic one. They exist in their respective restricted regions of the parameter space because of the limitations for all fractions to stay positive and do not exceed $1$. The linear stability analysis indicates the stability of both fixed points within their respective regions. Simple expressions obtained for the fractions $S^*$, $E^*$, $I^*$ and $R^*$ in the endemic fixed point enable to discuss the ways to bring the number of infected individuals in a stationary state down via changing the model parameters. This can be achieved by lowering the contact rate $\beta$ and/or by the increase of the identification $\alpha$ and isolation $\delta$ rates. However, if $\beta$ equals or exceeds the critical value $\beta_c$, the disease-free fixed point can not be achieved at any combination of $\alpha$ and $\delta$. This justifies the need for lowering the contact rate $\beta$ (quarantine measures) to control the dissemination of COVID-19.

Analytic solution of the $SEIRS$ is not possible, therefore, we employed numeric solution to examine the dynamics of the disease dissemination at various sets of model parameters. Numeric solution provides an evidence that, during the early-time evolution, the $R$ fraction is linearly proportional to the $E$ fraction. This simplifies the differential equations by bringing the model to the class of the $SIS$ model with rescaled contact rate $\beta'$ and enables an approximate analytic solution for the $SEIRS$ model at early times of its spread. Similar analysis is performed for the decay of the disease, given suitable choice of the model parameters. The exponential decay is found, governed solely by the curing rate $\gamma$. Both approximate solutions do agree well with the results of the numeric one, within their respective regions of validity. Both approximate solutions provide simple analytic expressions for the estimates of system dynamics at early stage and at its decay. The numeric solution can also be fitted well by a modified logistic function, as suggested in some previous works.

The effects of a quarantine and of its relaxation are modelled via step-wise switching between the ``normal life'' contact rate $\beta$ and that during a quarantine, $\beta_2<\beta$. The numeric solution can be used only in this case. The switch into a quarantine mode is performed if the fraction of newly identified infected individuals per unit time, $\dot{I}$ exceeds a quarantine threshold $\Delta$. At fixed values of $\alpha$ and $\delta$, the contact rate $\beta_{2,\mathrm{th}}$ exist, such that the disease decays only if the contact rate is lower than $\beta_{2,\mathrm{th}}$. It increases with the increase of the $\alpha$ and $\delta$ rates. Therefore, less strict quarantine measures are possible if more extensive testing of population and prompt isolation of infected individuals are undertaken. However, we also found that the optimal quarantine contact rate $\beta_2$ is about $0.8\beta_{2,\mathrm{th}}$ and further decrease of $\beta$ does not lead to faster decay of disease. This indicates no need for over-strong quarantine measures.

If, as the result of quarantine measures, $\dot{I}$ is reduced below $\epsilon$ (the quarantine relaxation threshold), then we restore the normal life contact rate $\beta$. Application of this algorithm of periodic switching on/off of a quarantine results in the wave-like behaviour for all the fractions of infected individuals. At fixed values of $\alpha$, $\delta$ and of both thresholds $\Delta$ and $\epsilon$, the waves height are determined by the quarantine threshold $\Delta$ only, whereas their separation in time -- by its relaxation threshold $\epsilon$. Therefore, one way to suppress the consequent waves of the disease in the $SEIRS$ model is: to fix identification $\alpha$ and isolation $\delta$ rates and to decrease $\Delta$ (undertake quarantine measures earlier) each time a quarantine is renewed. More realistic situation is, however, when both $\alpha$ and $\delta$ are small at the beginning of the pandemy (no testing algorithms, equipment and required chemicals available) but gradualy increase thereafter until the reasonable saturation for their values is reached. We modelled this scenario assuming quarantine on/off switch with fixed thresholds $\Delta$ and $\epsilon$ and the time-dependent $\alpha$ and $\delta$ rates using logistic function. Is is found that the heights of the second and of following waves of the disease dissemination in such scenario depend critically on the time scale of the growth of $\alpha$ and $\delta$. The dependence is rather non-linear, where a small increase of the time scale results in dramatic increase of the waves' height with the possibility of the second wave to be broader and higher than the first one. This shows the importance of the need for prompt introduction of testing and increasing the hospital level of readiness to avoid the wide and high second and following waves of the pandemy.

We did not adjust or fine-tune intentionally this modelling study for particular country/region/town and preferred to stand on rather general grounds. In this way the study concentrates on the general effects and disease behaviour patterns that are found at certain set of model parameters. It, of course, can be tuned for some special case given the relevant statistical data is available, but, in our view, the cellular automaton, geography-based modelling and network-based approaches suit this purpose much better. We see, however, the option for a synergy between the $SEIRS$ model developed here and above mentioned approaches and plan to do this in following studies.       

\section{Acknowledgements}

This work was supported by the National Research Foundation of Ukraine (Grant agreement No. 50/02.2020).
The computer simulations have been performed on the computing cluster of the Institute for Condensed Matter Physics of NAS of Ukraine (Lviv, Ukraine).

\bibliography{SEIR_covid.bib}{}
\end{document}